\documentclass[a4paper,12pt]{article}

\usepackage{amsmath,amssymb,amsfonts}
\usepackage[dvips]{graphicx}
\usepackage{epsfig}
\usepackage{bbding}
\usepackage{pifont}
\usepackage{wasysym}
\usepackage{tikz}
\usepackage{color}
\usepackage{amsmath}
\usepackage{amsfonts}
\usepackage{verbatim}
\usepackage{amssymb}
\usepackage{graphicx}
\usepackage{amsmath,amssymb,calc}
\usepackage{bbm}
\usepackage{setspace}
\usepackage{color}
\usepackage{amsmath}
\usepackage{amsfonts}
\usepackage{verbatim}
\usepackage{amssymb}
\usepackage{graphicx}
\usepackage{array}
\usepackage{epstopdf}

\input{my_opt.sty}

\usepackage{amsmath,amssymb,amsfonts,graphicx}
\usepackage{epsfig}
\usepackage[left=2.4cm,top=3.3cm,right=2.4cm,bottom=3.3cm,bindingoffset=0cm]{geometry}

\usepackage{tabularx}

\def\be{\begin{equation}}
\def\ee{\end{equation}}
\def\bqa{\begin{eqnarray}}
\def\bea{\begin{eqnarray}}
\def\eea{\end{eqnarray}}
\def\beq{\begin{equation}}
\def\eeq{\end{equation}}

\def\brc{\langle}
\def\ckt{\rangle}

\def\de{\partial}

\def\1{\mathbbm{1}}

\def\ta{\widetilde{a}}

\def\ta{\widetilde{a}}

\numberwithin{equation}{section}

\begin{document}

\title{
\begin{flushright}\ \vskip -1.5cm {\small {IFUP-TH-2019}}\end{flushright}
\vskip 20pt
\bf{ Dynamics  and symmetries  in \\  chiral $SU(N)$ gauge theories}
}
\vskip 30pt  
\author{  Stefano Bolognesi and
 Kenichi Konishi    \\[13pt]
{\em \footnotesize
Department of Physics ``E. Fermi", University of Pisa}\\[-5pt]
{\em \footnotesize
Largo Pontecorvo, 3, Ed. C, 56127 Pisa, Italy}\\[2pt]
{\em \footnotesize
INFN, Sezione di Pisa,    
Largo Pontecorvo, 3, Ed. C, 56127 Pisa, Italy}\\[2pt]
{ \footnotesize  stefano.bolognesi@unipi.it, \ \  kenichi.konishi@unipi.it}  
}

\vskip 6pt
\date{June  2019}
\maketitle
\vskip 0pt

\begin{abstract}

Dynamics and symmetry realization in various chiral gauge theories in four dimensions are investigated, 
generalizing a recent work by M. Shifman and the present authors \cite{BKS},  by relying on the standard 't Hooft anomaly matching conditions and on some other general ideas. 
These requirements are so strong that the dynamics of the systems are severely constrained. 
Color-flavor or color-flavor-flavor locking, dynamical Abelianization,  and combinations of these,  are  powerful ideas    which often leads to solutions of the anomaly matching conditions. 
Moreover, a conjecture is made on generation of a  mass hierarchy associated 
 with symmetry breaking in chiral gauge theories,  which has no analogues in vector-like gauge theories such as QCD. 

\end{abstract}
\newpage
\tableofcontents

\section{Introduction}

Our world has a nontrivial chiral structure. The macroscopic structures such as biological bodies
often have approximately left-right symmetric forms, but not exactly. At the molecular levels,  $O(10^{-6}   {\rm cm})$, the structure of DNA has 
a definite chiral spiral form.  At the microscopic length scales of the fundamental interactions, $O(10^{-14} {\rm cm})$,
 the left- and right-handed quarks and leptons have different couplings to the 
 $SU(3)\times SU_L(2)\times U_Y(1)$ gauge bosons. 
   In spite of the impressive success of the standard   model, 
 and after many years of theoretical 
 studies of four dimensional gauge theories,  our understanding of {\it strongly-coupled chiral}  gauge theories is today surprisingly limited\footnote{See however 
 \cite{Raby}-\cite{ShiShr} for partial list of earlier studies of these theories.  See \cite{Ryttov:2017gtm} for a recent work on the infrared fixed point in a class of chiral gauge theories.}.  
 An almost half-century of studies of vector-like gauge theories like $SU(3)$ quantum chromodynamics (QCD), based on 
lattice simulations with ever more powerful computers,  and roughly $\sim$ 25 years of beautiful theoretical developments in models with ${\cal N}=2$ supersymmetries, 
both concern vector-like theories only. 
  Perhaps it is not senseless to make some more efforts to understand this class of gauge theories, which
 Nature might be making use of, in an as yet unknown way to us.

Urged by such a motivation we have recently revisited the physics of some chiral gauge theories \cite{BKS}.   In the present paper we generalize the analysis done  there  to a wider class of models, and try to learn some general lessons from them. We use as guiding light the standard 't Hooft anomaly matching conditions \cite{th}.
  To be concrete, we shall limit ourselves to $SU(N)$ gauge theories with a set of Weyl fermions in a complex representation of the gauge group. 
 Also only asymptotically free type of models will be considered, as  weakly coupled infrared-free  theories can be reliably analyzed in perturbation theory, 
 as in the case of the standard electroweak model. 
For  simplicity  we shall restrict ourselves to various irreducibly chiral\footnote{For example we do not consider addition of fundamental-antifundamental pairs of  fermions. Models of this type, in the  simplest cases $(N_{\psi},N_{\chi}) = (1,0), (0,1) $, have been studied in \cite{ADS}. 
} 
 $SU(N)$   theories,    with $N_{\psi}$ fermions  $\psi^{\{ij\}}$  in the symmetric representation,    $N_{\chi}$  fermions $\chi_{[ij]}$ in the anti-antisymmetric representation,    and a number of anti-fundamental (or fundamental) multiplets, 
$\eta^a_i$  (or ${\tilde \eta}^{a\, i}$). The number of the latter is fixed by the condition that the gauge group be anomaly free.

\begin{figure}[h!t]
\begin{center}
\includegraphics[scale=0.8]{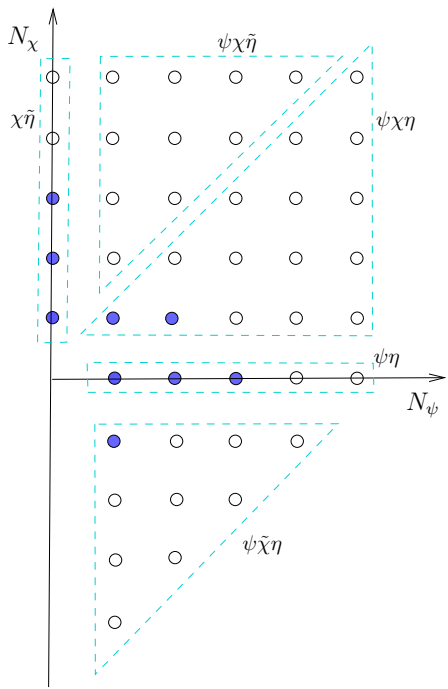}
\end{center}
\caption{\small A class of chiral QCD theories at large-$N$ in the plane $(N_{\psi},N_{\chi})$. } 
\label{theories}
\end{figure}

 Figure \ref{theories} gives a schematic representation of the various irreducibly $SU(N)$  chiral theories we shall be interested in.  Both $N_{\psi}$ and $N_{\chi}$ can go up to $5$ without loss of asymptotic freedom for large $N$. The ones we will explicitly consider are summarized in Table  \ref{tt} with their $b_0$ coefficient.
\begin{table}[h!t]
  \centering 
  \begin{tabular}{|c |c |  }
\hline
 Model  &  $ 3 b_0  $     \\
 \hline
  $(1,1) $   &   $9 N-8  $     \\
 $ (1,0)$      &   $  9N - 6  $     \\
  $ (2,0)$      &   $  7N - 12  $     \\
   $ (3,0)$      &   $  5N - 18  $     \\
    $ (0,1)$      &   $  9N + 6  $     \\
     $ (0,2)$      &   $  7N +12   $     \\
      $ (0,3)$      &   $  5N + 18  $     \\
       $ (2,1)$      &   $  7N - 14  $     \\
        $ (1,-1)$      &   $  7 N   $     \\
   \hline
\end{tabular}  
  \caption{First coefficients of the beta function.}
\label{tt}
\end{table}
 The gauge interactions in these models become strongly coupled in the infrared.  There are no gauge-invariant bifermion condensates,  no mass  terms or potential terms  (of renormalizable type) can be added to deform the theories, no $\theta$ parameter exists.  The main question we would like to address ourselves, given a model of this sort, is how to solve the 't Hooft anomaly matching conditions in the IR and if there are more than one apparently possible dynamical scenarios, all consistent with the matching conditions.

The paper is organized as follows. In Section~\ref{due} we revisit the $(N_{\psi},N_{\chi}) = (1,1)$ model previously considered in \cite{BKS}. In Sections \ref{tre}\,--\,\ref{dieci} we consider respectively the models  $(N_{\psi},N_{\chi}) = (1,0),(2,0),(3,0),(0,1),(0,2),(0,3),(2,1),(1,-1)$. In Section~\ref{undici} we discuss the pion decay constant and a possible new hierarchy mechanism. We conclude in Section~\ref{undici} trying to draw some general lesson for strongly-coupled chiral gauge theories. 
 Consistency check of the many proposed phases with the a-theorem and with the ACS  criterion  is done in Appendix~\ref{athACS}.

\section{Revisiting the $(N_{\psi},N_{\chi}) = (1,1)$   (``$\psi\chi\eta$")  model     }
\label{due}

We first review the analysis of the model with left-handed fermion matter fields  
\be   \psi^{\{ij\}}\;, \qquad  \chi_{[ij]}\;, \qquad    \eta_i^A\;,\qquad  A=1,2,\ldots 8\;,  
\ee
a symmetric tensor,  an  anti-antisymmetric tensor and eight  anti-fundamental multiplets  of $SU(N)$,
and add a few new comments with respect to \cite{BKS}.\footnote{Earlier studies on this model can be found in \cite{Goity:1985tf,Eichten:1985fs,AS}.}  It  is  asymptotically free, the first coefficient of the beta function being,
\be   b_0  = \frac 13\left[ 11N-  (N+2) -(  N-2) -8  \right] =  \frac{ 9 N- 8 }{3}\;.      \label{beta0}
\ee
 
 It is a very strongly coupled theory in the infrared and unlikely to flow into an infrared-fixed point CFT.  
A nonvanishing instanton amplitude 
\be  \brc \psi \psi \ldots  \psi \chi \chi \ldots  \chi  \eta ... \eta  \ckt \ne 0 
\ee
involves $N+2$ $\psi$'s, $N-2$ $\chi$'s and $8$ $\eta$'s.

The model has a global $SU(8)$ symmetry.  It has also three $U(1)$ symmetries,   $U_{\psi}(1)$,   $U_{\chi}(1)$, $U_{\eta}(1)$, of which two  combinations are anomaly-free. 
They can be taken e.g.,  as 
\bea &&   U_1(1): \qquad   \psi\to  e^{i  \frac{\alpha}{N+2}} \psi\;, \qquad \eta \to   e^{ -i \frac{\alpha}{8}} \eta\;;   
\nonumber \\
&&   U_2(1): \qquad   \psi\to  e^{i  \frac{\beta}{N+2}} \psi\;, \qquad \chi \to   e^{- i \frac{\beta}{N-2}} \chi\;.  \label{nonanU1}
\eea

There are also  anomaly-free  discrete subgroups  ${\mathbbm Z}_{N+2}\otimes {\mathbbm Z}_{N-2}\otimes  {\mathbbm Z}_{8}$   of $U_{\psi}(1)$,   $U_{\chi}(1)$, $U_{\eta}(1)$,
which are not broken by the instantons.  However, they are not independent of each other, in view of the nonanomalous symmetries  (\ref{nonanU1})
 The global continuous symmetry of the $\psi-\chi-\eta$ model is
\be    G_{\rm f}= SU(8) \times U_1(1)\times U_2(1)\;.
\ee

 \subsection{  Partial color-flavor locking} 
 
 Possible dynamical scenarios in this model have been analyzed and discussed in \cite{BKS}. 
 It was proposed that a possible phase (valid for $N \ge 12$)
  can be described by the nonvanishing bi-fermion condensates  
 \be   \langle  \phi^{i A} \rangle = \brc  \psi^{ij}  \eta_j^A  \ckt \;, \qquad    \langle  \tilde{\phi}^{i}_j \rangle \equiv  \langle  \psi^{ik} \chi_{kj}   \rangle\;.
\label{condensates}   \ee
 More concretely,  the proper realization of the global $SU(8)$ symmetry has led us  to assume the following form for these condensates:
 \beq
 \brc  \psi^{ij}  \eta_j^A  \ckt= \Lambda^3 \left(\begin{array}{c} c  {\mathbf 1}_{8}  \\ \hline
  \\
{\mathbf 0}_{N-8,8}\\
\\
\end{array}\right)^{i  \, A} \;, \qquad 
 \langle  \psi^{ik} \chi_{kj}   \rangle = \Lambda^3   \left(\begin{array}{c|ccc|c}
 a \,  {\mathbf 1}_{8}   &  &  & &  \\ \hline  &  d_1  &  & & \\ &   & \ddots &  & \\ &  &  & d_{N-12}&
\\ \hline & &  &&   b  \,  {\mathbf 1}_{4}   \end{array}\right)^i_j \;,   \label{conden}
\eeq
where 
\be    8 a + \sum_{i=1}^{N-12}  d_i + 4  b =0\;, 
\qquad     a, d_i, b \sim    O(1)\;. 
\ee
The symmetry breaking pattern is, therefore,
\beq
SU(N)_{\rm c} \times SU(8)_{\rm f}   \times U(1)^2  \to   SU(8)_{\rm cf} \times  U(1)^{N-11} \times SU(4)_{\rm c}\,.
\label{scenario1}
\eeq
 The theory dynamically Abelianizes (in part). $SU(8) \subset SU(N)$ is completely Higgsed but due to color-flavor (partial) locking no NG bosons appear  in this sector (the would-be NG bosons make the $SU(8) \subset SU(N)$  gauge bosons massive.) Only $SU(4) \subset SU(N)$ remains unbroken and confining. 
The remainder of the gauge group Abelianizes.   The baryons 
\be {\tilde B}^A_j =   \psi^{ik}  \chi_{[kj]} \eta_i^A \sim  \eta_j^A\;,    \qquad   (9 \le j \le N-4) \ee and     
\be  B^{\{AB \}}=    \psi^{ij}  \eta_i^A \eta_j^B,  \ee 
symmetric in the flavor indices $(A\leftrightarrow B)$,\footnote{If the massless  $B^{\{AB \}}$ were antisymmetric in the flavor indices, they would contribute $8-4=4$ to the $SU(8)$ anomaly.  We would then need $N-4$  massless fermions of the form  ${\tilde B}^A_j \sim  \eta_j^A$,
but this is impossible as the latter arises from the Abelianization of the rest of the color gauge group,  $SU(N-8)$. }   remain massless  and together saturate the  't Hooft  anomaly matching condition for $SU(8)$:  
\be    8+4 +  N-12 =N\;. 
\ee
Note that both nonanomalous continuous $U_{1,2}(1)$'s are broken by the two condensates. 
Actually, for some $N$, a discrete symmetry survives the condensates of the form (\ref{condensates}), and  the discrete anomaly matching must be taken into account.

\subsubsection{Discrete symmetries}

Under the discrete symmetries the fields transform as 
\bea && {\mathbbm Z}_{N+2} \subset  U_{\psi}(1): \qquad  \psi\to  e^{i \tfrac{2\pi k}{N+2}} \psi\;,     \quad k=0, 1,\ldots  N+1\;;
\nonumber \\
&&{\mathbbm Z}_{N-2} \subset  U_{\chi}(1): \qquad  \chi  \to  e^{i \tfrac{2\pi \ell }{N-2}} \chi\;,    \quad \ell=0, 1,\ldots  N-3\;;
\nonumber \\
   &&\ \ \ \,  {\mathbbm Z}_{8} \subset  U_{\eta}(1): \qquad    \, \eta  \to  e^{i  \tfrac{2\pi m}{8} } \eta  \;, \quad m=0, 1,\ldots  7\;.\label{discrete}
\eea
A discrete  subgroup   survives the condensates 
(\ref{condensates}) if
\be      \frac{k}{N+2}  -  \frac{\ell}{N-2} \in {\mathbbm Z}\;,\qquad  \frac{k}{N+2}  - \frac{m}{8}  \in {\mathbbm Z}\;.
\ee
Clearly there are no discrete surviving symmetry for odd $N$. For $N$ even, the above shows that there remains a ${\mathbbm Z}_2$ symmetry, for $N= 4 n$, $n \in {\mathbbm Z}$, or a  ${\mathbbm Z}_4$ symmetry, for $N= 4 n +2$.

To be concrete, consider $N=14$.  The conditions above read in this case
\be   \frac{k}{16}  -  \frac{\ell}{12} \in {\mathbbm Z}\;,\qquad  \frac{k}{16}  - \frac{m}{8}  \in {\mathbbm Z}\;.
\ee
The transformation
\be   \psi \to  e^{{ \pi i}/ {2}} \psi\;, \qquad  \chi \to  e^{-\pi i /2} \chi\;, \qquad  \eta \to  e^{-\pi i /2} \eta \;, 
\ee
generates  ${\mathbbm Z}_4$, which is kept unbroken by $\brc \psi \chi \ckt$ and $\brc \psi \eta \ckt$.  The ${\mathbbm Z}_4$ charge 
of $(\psi, \chi, \eta)$ fields are $(1, -1, -1)\ {\rm Mod}\ 4$.

Consider the discrete  anomaly $SU(8)^2 \,   {\mathbbm Z}_4$ \cite{Csaki:1997aw}.  In the UV,  the only contribution is from the $\eta$ fields, which gives
\be      N \cdot  1 \cdot (-1)=- N= - 14\;.
\ee
In the IR,  $\eta_j^A  \,   (9 \le j \le N-4) $  gives
\be    (N-12)  \cdot   1 \cdot  (-1)   =  -2\;,
\ee
whereas $B^{\{ AB \}}=    \psi^{ij}  \eta_i^A \eta_j^B$   contribute
\be    1   \cdot     (8+2) \cdot  (-1) = -10\;,  
\ee
total  of 
\be    -2-10=-12\;.  
\ee
The difference between UV and IR is 
\be     -14-(-12) = -2  \ne  0\;  \quad  \,{\rm Mod}\,\,  4\;.  
\ee
Thus the discrete  $SU(8)^2 \,   {\mathbbm Z}_4$  anomaly does not match for $N=14$. 
A similar situation is found for all  $N$ of the form $ 4 n +2$, $n =3,4,5,\ldots$.  

As for the discrete ${\rm Grav}^2 \, {\mathbbm Z}_4$ anomaly,  we count only the ${\mathbbm Z}_4$ charges and the multiplicities:
in the UV, it is
\be      N \cdot  1 \cdot (-1)= N= - 14\;,
\ee
whereas in the IR the value is 
\be    2 \cdot   1 \cdot  (-1)   +   \frac{8\cdot 9}{2} \cdot (-1)=  -38\;.
\ee
The difference is 
\be   38-14 = 24 = 0\, \quad  \,  {\rm Mod}\, \ 4\;,
\ee
so it is matched.

For $N= 4 n$,    the conditions
\be      \frac{k}{4 n+2}  -  \frac{\ell}{4 n-2} \in {\mathbbm Z}\;,\qquad  \frac{k}{ 4 n +2}  - \frac{m}{8}  \in {\mathbbm Z}\;.
\ee
leaves a ${\mathbbm Z}_2$ symmetry generated by the transformations with $k=2n+1$, $\ell=2n-1$ and $m=4$\;.  It is easy to verify that all the discrete anomalies involving ${\mathbbm Z}_2$  are matched in the UV and in the IR.

The fact that the discrete anomaly matching does not work for $N= 4 n +2$ renders the scenario  (\ref{condensates})-(\ref{scenario1})
not  likely to be realized for any $N$.   There are however other possibilities as discussed below.

\subsection{Color-flavor locking and dynamical Abelianization: an alternative scenario  \label{altern1}}

Another possible phase, for $N \ge 8$,  which was not considered in \cite{BKS},    is described by the condensates (\ref{condensates}), but  this time of the form
 \be
  \brc  \psi^{ij}  \eta_j^A  \ckt    = \Lambda^3 \left(\begin{array}{c} c  {\mathbf 1}_{8}  \\ \hline
  \\
{\mathbf 0}_{N-8,8}\\
\\
\end{array}\right)^{i  \, A}  
 \;, \qquad 
\langle  \psi^{ik} \chi_{kj}   \rangle  = \Lambda^3   \left(\begin{array}{c|ccc}
   {\mathbf 0 }_{8}   &  &   & \\
    \hline 
     &  d_1  &  &  \\  &   & \ddots   &   \\ &  &  &    d_{N-8}
  \end{array}\right)^i_j
  \;.  \label{condenBis}
\ee
The symmetry breaking pattern is:  
\be     SU(N)\times SU(8)\times U(1)^2  \to   SU(8)_{\rm cf} \times U(1)^{N-8}\;.   \label{scenario1bisbis}
\ee
As  $U(1)^{N-8}$ is an Abelian subgroup of the color $SU(N)$, whereas both nonanomalous flavor   $U(1)$ are broken by the condensates,   we shall 
consider only the  $  SU(8)_{\rm cf}^3$  anomalies.   Indicating the color indices up to $8$ by  $i_1$ or $j_1$  while those larger than $8$ by 
$i_2$ or $j_2$,  one has the decomposition of the fields  in  $SU(8)_{\rm cf} $ multiplets, see Table~\ref{Simplest002}.   
\begin{table}[h!]
  \centering 
  \begin{tabular}{|c|c|c |}
\hline
$ \phantom{{{   {  {\yng(1)}}}}}\!  \! \! \! \! \!\!\!$   & fields         &    $ SU(8)_{\rm cf} $         \\
 \hline
   $ \phantom{{\bar{ \bar  {\bar  {\yng(1)}}}}}\!  \! \! \! \! \! \!\!\!$  {\rm UV}&   $\psi^{i_1 j_1}  $      &    $  {\bar { \yng(2) }}  $      \\
&    $\psi^{i_1 j_2}  $      &    $  (N-8)\cdot  {\bar { \yng(1) }}  $         \\
 &     $\psi^{i_2 j_2}  $      &    $ \frac{(N-8)(N-7)}{2} \cdot (\cdot)    $        \\
 &   $\chi_{i_1, j_1}  $      &    $  \yng(1,1)  $         \\
 &     $\chi_{i_1, j_2}  $      &    $  (N-8)\cdot    \yng(1)  $         \\
 & $\chi_{i_2, j_2}  $      &    $ \frac{(N-8)(N-9)}{2} \cdot (\cdot)   $        \\
 & $ \eta^{A}_{j_1}$      &   $        \yng(1) \otimes \yng(1)=  \yng(2) \oplus \yng(1,1)$          \\
&  $ \eta^{A}_{j_2}$      &   $     (N-8)\cdot     \yng(1)$     \\
   \hline 
\phantom{\huge i}$ \! \!\!\!${\rm IR }&  $  {\tilde B}^A_{j_2}  \sim  \eta^{A}_{j_2} $      &   $     (N-8)\cdot     \yng(1)$       \\
&  $  B^{[A j_1]}   \sim  {\cal A} (\eta^{A}_{j_1})    $      &   $ \yng(1,1)$        \\    
&  $  {\hat  B}^{[i_1 j_1]}      \sim  \chi_{i_1 j_1}  $      &    $  \yng(1,1)  $       \\
\hline
\end{tabular}
  \caption{ \footnotesize  $B^{[AB]}  \sim    \psi^{ij}  \eta^{[A}_i  \eta^{B]}_j$ and    $ {\hat  B}^{[AB]}  \sim  (\psi \eta)^{A,i}    \chi_{ij}     (\psi \eta)^{B, j} $.    
  $ \eta^{A}_{j_2} $  are weakly coupled due to the Abelianization of the  $SU(N-8)\times U(1) \subset  SU(N). $   They can be interpreted as 
  $ {\tilde B}^A_{j} \sim    (\psi \chi \eta)^A_j$.   
The color indices up to $8$ are indicated by  $i_1$ or $j_1$  while those larger than $8$ by $i_2$ or $j_2$. 
}
  \label{Simplest002}
\end{table}
The massless baryons are shown in the lower part of the  Table~\ref{Simplest002}.  The $SU(8)^3$  matching works, as  in the infrared, 
\be   (N-8) + (8-4) + (8-4)   =N\;.
\ee
As for the discrete symmetry,  the surviving symmetry is  either ${\mathbbm Z}_2$, for $N= 4 n$, $n \in {\mathbbm Z}$, or ${\mathbbm Z}_4$ symmetry, for $N= 4 n +2$ under which the fields $\psi,\, \chi, \,\eta$ are charged with $(1,-1,-1)$.    An inspection of Table~\ref{Simplest002} shows that 
all discrete anomaly matching is also satisfied in this case, in contrast to the previous case.

\subsection{Partial color-flavor locking for $N\le 8$  \label{altern2}}

For $N <8$ the scenario above is not viable.  It is possible however that the color-flavor locking still takes place in
a different way (this possibility was not considered in \cite{BKS} either).  Let us assume that 
\be \brc  \psi^{ij}  \eta_j^A  \ckt   = \Lambda^3 \Big(\begin{array}{c|c} c  {\mathbf 1}_{N} &
{\mathbf 0}_{N,8-N}
\\
\end{array}\Big)^{i  \, A}  
 \;, \qquad 
\langle  \psi^{ik} \chi_{kj}   \rangle  =0
  \;. 
\ee
The symmetry  breaking pattern is now
\be     SU(N)\times SU(8)\times U(1)^2  \to   SU(N)_{\rm cf} \times  SU(8-N) \times  {\tilde U}(1)\;.   \label{scenario1tris}
\ee
The fermions decompose as  in Table~\ref{ColorFlavortris}.
\begin{table}[h!t]
  \centering 
  \begin{tabular}{|c|c|c |c|c|  }
\hline
$ \phantom{{{   {  {\yng(1)}}}}}\!  \! \! \! \! \!\!\!$   & fields   &  $SU(N)_{\rm c}  $    &  $ SU(8-N)$     &   $ {\tilde U}(1)   $  \\
 \hline
    $ \phantom{{\bar{ \bar  {\bar  {\yng(1)}}}}}\!  \! \! \! \! \! \!\!\!$   {\rm UV}&   $\psi$   &   $ {\bar { \yng(2)}} $  &    $  \frac{N(N+1)}{2} \cdot (\cdot) $    & $N+2$    \\
&$\chi$   &   $ { \yng(1,1)} $  &    $  \frac{N(N-1)}{2} \cdot (\cdot) $    & $-  \frac{(N-6)(N+2)}{N-2}$    \\
 & $ \eta^{A_1}$      &   $  \yng(2) +  \yng(1,1)$     & $N^2 \, \cdot  \, (\cdot) $     &   $ - (N+2)  $ \\
 &  $ \eta^{A_2}$      &   $ (8-N) \cdot \yng(1)$     & $N \, \cdot  \,  \yng (1) $     &   $ - (N+2)  $ \\
  \hline
  $ \phantom{{\bar{   {  {\yng(1,1)}}}}}\!  \! \! \! \! \!\!\!$  {\rm IR}&   $ B^{[A_1 B_1]  }$      &   $  \yng(1,1)$     & $\frac{N(N-1)}{2}   \, \cdot  \, (\cdot) $     &   $ - (N+2)  $ \\
   &   $ B^{[A_1  B_2 ]  }$      &   $ (8-N) \cdot \yng(1)$     & $N \, \cdot  \,  \yng (1) $     &   $ - (N+2)  $ \\
   &     $ {\hat B}^{[A_1 B_1]}$      &   $\yng(1,1)$     & $\frac{N(N-1)}{2}   \, \cdot  \, (\cdot) $     &   $ -  \frac{(N-6)(N+2)}{N-2} $ \\
        \hline
\end{tabular}
  \caption{\footnotesize  Partial color-flavor locking for $N\le 8$ and the $SU(8)$ anomaly matching of Subsection \ref{altern2}.     
  $A_1, B_1$ stand for the flavor indices up to $N (<8)$,  $A_2, B_2$  for the rest. }
  \label{ColorFlavortris}
\end{table}
The massless baryons which saturate the anomalies are made of
\bea  &&B^{[A_1  B_1]  } =\psi^{ij}  \eta_i^{A_1} \eta_j^{B_1} \sim     \yng(1,1)  \;,  \qquad B^{[A_1  B_2 ]  } =\psi^{ij}  \eta_i^{A_1} \eta_j^{B_2}  \sim    \eta^{B_2}   \;,  \nonumber \\
      \phantom{ \bar {\yng(1)} }&& \ \ \  \qquad \qquad \quad  {\hat B}^{[AB]} =   \psi^{ik} \eta_k^A  \chi_{ij}  \psi^{j \ell} \eta_{\ell}^B \sim    \chi_{AB}\;.\ 
\eea

\subsection{Full Abelianization and general $N$
\label{fulla}}

The dynamical scenarios (\ref{scenario1})  assumes that $N \ge  12$, whereas the one in  (\ref{scenario1bisbis}) 
requires $N\ge 8$ and  (\ref{scenario1tris}) requires  $N\le 8$.

Still another option,  consistent for any value of  $N$,  considered  in \cite{BKS},     is that 
the gauge group dynamically Abelianizes completely, by the adjoint condensates
\be   
  \brc  \psi^{ij}  \eta_j^A  \ckt    = 0
 \;, \qquad 
\langle  \psi^{ik} \chi_{kj}   \rangle  = \Lambda^3   \left(\begin{array}{ccc}
 
      d_1  &  &  \\     & \ddots   &   \\   &  &    d_{N}
  \end{array}\right)^i_j
  \;. 
\ee
with $  \sum_j  d_j=0$ and no other particular relations among $d_j$'s. No color-flavor locking takes place.
The symmetry breaking occurs as:
  \beq     
  SU(N)_{\rm c} \times SU(8)_{\rm f} \times U(1)^2 \to \prod_{\ell=1}^{N-1}  U_{\ell} (1) \times  SU(8)_{\rm f} \times {\tilde U}(1)\,,
  \label{scenario3}
  \eeq
  where ${\tilde U}(1)$ is an unbroken combination of the two nonanomalous  $U(1)$'s,  (\ref{nonanU1}), with charges:
   \be       \psi:   2 \;,\qquad   \chi:   - 2\;,\qquad    \eta:   -1  \;.
  \ee
The fields  $\eta^A_i$ are all massless and  weakly coupled (only to the gauge bosons from the Cartan subalgebra which we will refer to as the photons; they are infrared free) in the infrared. Also, some of the fermions $\psi^{ij}$  do not participate in the condensates. Due to the fact that 
$\psi^{\{ij\}}$ are symmetric whereas $\chi_{[ij]}$ are antisymmetric,  actually only nondiagonal elements of   $\psi^{\{ij\}}$  condense and get mass. The diagonal 
fields $\psi^{\{ii\}}$, $i=1,2,\ldots, N$ remain massless and weakly coupled.   Also there is one NG boson. 
\begin{table}[h!t]
  \centering 
  \begin{tabular}{|c|c|c |c|  }
\hline
$ \phantom{{{   {  {\yng(1)}}}}}\!  \! \! \! \! \!\!\!$   & fields          &  $ SU(8)$     &   $ {\tilde U}(1)   $  \\
 \hline
  \phantom{\huge i}$ \! \!\!\!\!$  {\rm UV}& $\psi$      &    $   \frac{N(N+1)}{2} \cdot  (\cdot) $    & $  \frac{N(N+1)}{2} \cdot (2)$    \\
  &  $\chi$      &    $   \frac{N(N-1)}{2} \cdot  (\cdot) $    & $  \frac{N(N-1)}{2} \cdot (-2)$        \\
 &$ \eta^{A}$      &   $ N \, \cdot  \, {\yng(1)}  $     &   $  8N \cdot  (-1) $ \\
   \hline 
  \phantom{\huge i}$ \! \!\!\!\!$  {\rm IR}&       $ (\psi \chi \psi)^{ii} \sim \psi^{ii}  $      &  $ N \cdot ( \cdot)   $        &    $  N \cdot (2) $   \\
     &  $ \psi \chi \eta^A  \sim   \eta^{A} $      &  $ N \, \cdot  \, {\yng(1)}  $        &    $  8  N \cdot (-1) $   \\
\hline
\end{tabular}  
  \caption{\footnotesize  Full dynamical Abelianization in the $\psi\chi\eta$ model, in Subsection~\ref{fulla}}
\label{Simplest}
\end{table}
The anomaly matching works as shown in Table \ref{Simplest}.

\section{$(N_{\psi},N_{\chi}) = (1,0)$ }
\label{tre}

Let us  review the $(N_{\psi},N_{\chi}) = (1,0)$ model studied in \cite{Dimopoulos:1980hn,ACSS,ADS,Poppitz,ShiShr}.
The matter fermions are
\beq
   \psi^{\{ij\}}\,, \quad    \eta_i^B\, , \qquad \;  B=1,2,\ldots , N+4\,,
\eeq
or
\be       \yng(2) +   (N+4) \, {\bar {\yng(1)}}\;. 
\ee
   The first coefficient of the beta function is
\be   b_0=\frac 13\left[ 11N -  (N+2) - (N+4) \right]= \frac{9N-6}{3}\;.
\ee
The (continuous) symmetry of this model is 
\beq
SU(N)_{\rm c} \times  SU(N+4)_{\rm f} \times {U}(1)\,,  \label{full}
    \eeq
    where $U(1)$ is an anomaly-free combination of  $U_{\psi}(1)$ and $U_{\eta}(1)$, with 
    \be     Q_{\psi}:\quad   N+4 \;, \qquad  Q_{\eta}:\quad    -(N+2) \;.  \label{comb}
    \ee
    There are also discrete symmetries 
    \be    {\mathbbm Z}_{\psi}  =  {\mathbbm Z}_{N+2}  \subset  U_{\psi}(1)\;, \qquad    {\mathbbm Z}_{\eta}  ={\mathbbm Z}_{N+4}  \subset  U_{\eta}(1)\;. 
    \ee

    \subsection{Chirally symmetric phase   in the $(1,0)$ model
\label{possible0}} 
    
    Let us first examine the possibility that no condensates form, the system confines and the flavor symmetry is unbroken \cite{Dimopoulos:1980hn}.
    The candidate massless baryons are:
      \be     B^{[AB]}=    \psi^{ij}  \eta_i^A  \eta_j^B \;,\qquad  A,B=1,2, \ldots, N+4\;,
\ee
antisymmetric in  $A \leftrightarrow B$.
All the $SU(N+4)_{\rm f}\times U(1)$ anomalies are saturated by  $ B^{[AB]}$ as can be seen by inspection of  the Table~\ref{Simplest0}. 
The discrete anomaly   ${\mathbbm Z}_{\psi}\,  SU(N)^2$  is also matched, as can be easily checked. 
\begin{table}[h!t]
  \centering 
  \begin{tabular}{|c|c|c |c|c|  }
\hline
$ \phantom{{{   {  {\yng(1)}}}}}\!  \! \! \! \! \!\!\!$   & fields  &  $SU(N)_{\rm c}  $    &  $ SU(N+4)$     &   $ { U}(1)   $  \\
 \hline 
  \phantom{\huge i}$ \! \!\!\!\!$  {\rm UV}&  $\psi$   &   $ { \yng(2)} $  &    $  \frac{N(N+1)}{2} \cdot (\cdot) $    & $   N+4$    \\
 & $ \eta^{A}$      &   $  (N+4)  \cdot   {\bar  {\yng(1)}}   $     & $N \, \cdot  \, {\yng(1)}  $     &   $  - (N+2) $ \\
   \hline     
 $ \phantom{ {\bar{   { {\yng(1,1)}}}}}\!  \! \! \! \! \!\!\!$  {\rm IR}&    $ B^{[AB]}$      &  $  \frac{(N+4)(N+3)}{2} \cdot ( \cdot )    $         &  $ {\yng(1,1)}$        &    $ -N    $   \\
\hline
\end{tabular}
  \caption{\footnotesize  Chirally symmetric phase of the  $(1,0)$  model.}\label{Simplest0}
\end{table}

\subsection{Color-flavor locked Higgs phase   \label{possible0bis}}

It is also possible that a color-flavor locked phase appears \cite{ADS,BKS}, with 
\be    \brc  \psi^{\{ij\}}   \eta_i^B \ckt =\,   c \,  \Lambda^3   \delta^{j B}\;,   \qquad   j, B=1,2,\dots  N\;,    \label{cflocking}
\ee
in which the symmetry is  reduced to 
\beq
SU(N)_{\rm cf} \times  SU(4)_{\rm f}  \times U^{\prime}(1) \,.
    \eeq
    As this forms a subgroup of  the full symmetry group, (\ref{full}),  it is quite easily seen, by making the decomposition of the fields in the direct sum of representations in  the subgroup,  that a subset of the same baryons 
    saturate all of the triangles associated with the reduced symmetry group, see Table~\ref{SimplestBis}.  
\begin{table}[h!t]
  \centering 
  \begin{tabular}{|c|c|c |c|c|  }
\hline
$ \phantom{{{   {  {\yng(1)}}}}}\!  \! \! \! \! \!\!\!$   & fields   &  $SU(N)_{\rm cf}  $    &  $ SU(4)_{\rm f}$     &   $  U^{\prime}(1)   $  \\
 \hline
   \phantom{\huge i}$ \! \!\!\!\!$  {\rm UV}&  $\psi$   &   $ { \yng(2)} $  &    $  \frac{N(N+1)}{2} \cdot   (\cdot) $    & $   1  $    \\
 & $ \eta^{A_1}$      &   $  {\bar  {\yng(2)}} \oplus {\bar  {\yng(1,1)}}  $     & $N^2 \, \cdot  \, (\cdot )  $     &   $ - 1 $ \\
&  $ \eta^{A_2}$      &   $ 4  \cdot   {\bar  {\yng(1)}}   $     & $N \, \cdot  \, {\yng(1)}  $     &   $ - \frac{1}{2}  $ \\
   \hline 
   $ \phantom{{\bar{ \bar  {\bar  {\yng(1,1)}}}}}\!  \! \!\! \! \!  \!\!\!$  {\rm IR}&      $ B^{[A_1  B_1]}$      &  $ {\bar  {\yng(1,1)}}   $         &  $  \frac{N(N-1)}{2} \cdot  (\cdot) $        &    $   -1 $   \\
       &   $ B^{[A_1 B_2]}$      &  $   4 \cdot {\bar  {\yng(1)}}   $         &  $N \, \cdot  \, {\yng(1)}  $        &    $ - \frac{1}{2}$   \\
\hline
\end{tabular}  
  \caption{\footnotesize   Color-flavor locked phase in the $(1,0)$ model, discussed in Subsection~\ref{possible0bis}.
  $A_1$ or $B_1$  stand for  $A,B=1,2,\ldots, N$,   $A_2$ or $B_2$ the rest of the flavor indices. 
   }\label{SimplestBis}
\end{table}

The discrete anomaly   ${\mathbbm Z}_{\psi} $ is broken by the condensate $\psi \eta$. There is (for generic $N$) no combination between 
${\mathbbm Z}_{\psi} $ and ${\mathbbm Z}_{\eta} $  which survives, therefore there is no discrete anomaly matching condition.

    It is not known which of the possibilities,    \ref{possible0} or \ref{possible0bis}, 
     is realized in the     $(1,0)$ model.     The low-energy degrees of freedom are  $\tfrac{(N+4)(N+3)}{2}$ massless baryons in  the former case,  
     and $\tfrac{N^2+7N}{2}$ massless baryons together with  $8N+1$  Nambu-Goldstone bosons, in the latter. 
     Thus the complementarity \cite{Fradkin}, as noted in \cite{BKS}, does not work here even though the (dynamical) Higgs scalars    $\psi \eta$  are in the fundamental representation of color.

\section{  $(N_{\psi},N_{\chi}) = (2,0)  $ } 
\label{quattro}

This is a straightforward generalization of the $\psi \eta$ model above. The matter fermions are
\beq
   \psi^{\{ij,\, m\}}\,, \quad    \eta_i^B\, , \qquad  m=1,2\;, \quad  B=1,2,\ldots , 2(N+4)\,,
\eeq
or
\be      2\,  \yng(2) +   2(N+4) \, {\bar {\yng(1)}}\;. 
\ee
The (continuous) symmetry of this model is 
\beq
SU(N)_{\rm c} \times SU(2)_{\rm f} \times SU(2N+8)_{\rm f} \times {U}(1)\,,
    \eeq
    where $U(1)$ is an anomaly-free combination of  $U_{\psi}(1)$ and $U_{\eta}(1)$,
    \be  U(1): \qquad      \psi \to e^{i \alpha / 2 (N+2)} \psi\;, \qquad  \eta \to  e^{-   i \alpha / 2 (N+4)} \eta\;.  \label{comb2}
    \ee
   The first coefficient of the beta function is
     \beq    
 b_0= \frac 13\left[11 N   -  2  (N+ 2)  -  2  (N+4) \right] =  \frac{ 7 N  -  12}{3}  \;,
\eeq
which is positive for  $N \ge 2$.

\subsection{No chiral symmetry breaking in the   $(2,0)$ model?   \label{impossible1}} 

  Let us first assume that no condensates form and no flavor symmetry breaking occurs.
  Assuming confinement, 
the possible massless baryons are
\be     B^{m, AB}=    \psi^{ij\,,m}  \eta_i^A  \eta_j^B \;.
\label{bar}
\ee
They cannot however saturate the triangles associated with the flavor symmetry 
\be  SU(2)_{\rm f} \times SU(2N+8)_{\rm f} \times {U}(1) \;.
\ee
For instance the  $ SU(2N+8)^3$ anomaly,  which is equal to $N$ in the UV, would be at least $\sim 2N$ for any baryon like (\ref{bar}) and thus it is not reproduced in any way in IR. We must conclude that confinement phase with unbroken flavor symmetries cannot be realized in this system. 
  This is in contrast to the $(1,0)$ model,  reviewed  in Subsection~\ref{possible0}.
    
\subsection{Partial color-flavor locking? \label{impossible3}} 

Let us consider next a partial color-flavor locking condensates 
\be    \brc  \psi^{\{ij\,, 1\}}   \eta_i^B \ckt = \, c \,\Lambda^3 \delta^{j B}\;,   \qquad   j, B=1,2,\dots  N\;,    \label{condens}
\ee
 which breaks   the symmetry to 
\be        SU(N)_{\rm cf} \times  SU(N + 8) \times {\tilde U}(1)\;;  \label{phase2}
\ee
$SU(2)$ is broken.  ${\tilde U}(1)$ is a linear combination of   (\ref{comb2}) and 
\be     U_1(1)=   \left(\begin{array}{cc}-\frac{1}{N}  {\mathbbm 1}_N & 0 \\0 & \frac{1}{N+8}  {\mathbbm 1}_{N+8}\end{array}\right)  \subset SU(2(N+4))\;.
\ee
So the unbroken ${\tilde U}(1)$  acts on the fields as
\bea   &&   \psi \to e^{i \tfrac{\alpha} { 2 (N+2)} } \psi\;;
\nonumber \\ &&\eta^A_i \to   e^{- i \tfrac{\alpha }{ 2 (N+2)}}    \eta^A_i\;, \qquad   (A=1,2, \ldots, N)\;;
\nonumber \\  &&  \eta^A_i \to   e^{-  i \tfrac{\alpha (N+4) }{ 2(N+2)(N+8)} }  \eta^A_i\;, \qquad   (A=N+1,N+2, \ldots, 2N+8)\;.
\eea
The charges with respect to ${\tilde U}(1)$ are: 
\be   \psi: \quad 1\;,\qquad \eta^<:  \quad  -1\;, \qquad  \eta^>:  \quad  -\frac{N+4}{N+8}\;.
\ee
The massless baryons are assumed to be of the form,
\beq   
B^{AB} =   \psi^{ij\;, 1}  \eta_i^A   \eta_j^B\;,   \qquad   A,B = 1,2,\ldots, N 
\label{ss39}
\eeq
and 
\beq  
\tilde{B}^{AB} =  \psi^{ij\;, 1 }  \eta_i^A   \eta_j^B\,,   \quad   A = 1,2,\ldots, N\,,  \quad   B=N+1, \ldots 2N+8\,.  
\label{ss40}
\eeq
Here we must choose 
$B^{AB}$  in the  symmetric  or antisymmetric representation of  the  $SU(N)_{\rm cf}$ group while  ${\tilde B}^{A B} $ is in the $({\underline N}, {\underline {N+8}})$ of  $SU(N)_{\rm cf} \times SU(N+8)_{\rm flavor}$.       ${\tilde U}(1)$  charges of   $B^{AB}$ and  $\tilde{B}^{AB} $ are
\be     B^{AB} :\; \quad  -1\;; \qquad     \tilde{B}^{AB}  :\; \quad  -\frac{N+4}{N+8}\;.
\ee
These assumptions are made such that the $SU(N+8)_{f}^3$ and  ${\tilde U}(1)  SU(N+8)_{\rm f}^2$ anomalies are matched in the UV and IR;  however, 
it is easy to verify that  the triangles   ${\tilde U}(1)^3$  and  $SU(N)_{\rm cf}^3$  cannot be matched. 
Therefore  the phase  (\ref{condens}), (\ref{phase2}),   cannot be realized.

\subsection{A possible phase: a double color-flavor  locking \label{possible2}} 
Another  possibility is to assume a double $SU(N)$ color-flavor-flavor locking
\bea && \brc  \psi^{\{ij\,, 1\}}   \eta_j^B \ckt = \, c \,\Lambda^3  \delta^{i,\,  B }\;,   \qquad \quad \     j, B=1,2,\dots  N\;,   
\nonumber \\   && \brc  \psi^{\{ij\,, 2\}}   \eta_j^B \ckt = \, c' \,\Lambda^3   \delta^{i,\,  B-N }\;,   \qquad   j=1,2,\dots  N\;, \quad B=N+1,\ldots, 2N \;,   \label{condens11}
\eea
The symmetry is broken to
\be  
 SU(N)_{\rm cf} \times  {\tilde U}(1) \times U^{\prime}(1)  \times  SU(8)\;.\label{thistime}
\ee
where ${\tilde U}(1)$ acts as before:
\be     \psi:    \quad 1  \;;\qquad \eta^{B\le 2N} :\quad -1\;;\qquad  \eta^{B> 2N} :\quad -\frac{1}{2}\;.
\ee 
There are 
\be     3 N^2  + 32 N + 3  
\ee
NG bosons.
$U^{\prime}(1)$ is a subgroup of  $SU(2)_{\rm ff} $  defined below,  (\ref{su2ff}), (\ref{su2fff})  which survives the condensates (\ref{condens11}).

In order to saturate all the anomalies,  one assumes that somehow only  
\beq  
\hat{B}^{A, B}=      \psi^{ij\;, 1}  \eta_i^A    \eta_j^B\,,   \qquad   A = 1,2,\ldots, N\,,    \quad   B=2N+1, \ldots 2N+8\,.  
\label{ss42Bis}
\eeq
or 
\beq  
\tilde{B}^{A, B}=      \psi^{ij\;, 1}  \eta_i^A    \eta_j^B\,,   \qquad   A = N+1,N+2,\ldots, 2N\,,    \quad   B=2N+1, \ldots 2N+8\,.  
\label{ss42Bisbi}
\eeq
(but not both)
remain massless.  One could write these states as 
\beq  
\hat{B}^{B}_{a} =    \sum_{A=1}^{2N} \sum_{m=1,2}  c_{a,  m, A}   \psi^{ij\;, m}  \eta_i^A   \eta_j^B\,,   \qquad   a = 1,2,\ldots, N\,,    \quad   B=2N+1, \ldots 2N+8\,.  
\label{ss42bbb}
\eeq
  Furthermore, we shall need also  
two types of baryons
\bea  && B^{[AB],\, 1} =    \psi^{ij\;, 1}  \eta_i^A   \eta_j^B\;,   \qquad   A,B = 1,2,\ldots, N  \ , \nonumber \\  && B^{[AB],\, 2} =    \psi^{ij\;, 1}  \eta_i^A   \eta_j^B\;,   \qquad   A,B = N+1,N+2,\ldots, 2N  \ ,
\eea
both antisymmetric in   $AB$,  all of them remaining  massless. 
It is a simple exercise to check that all anomalies,  
  $SU(8)^3$, $SU(8)^2 {\tilde U}(1)$, ${\tilde U}(1)^3$,  ${\tilde U}(1)$,  $SU(N)^3$,  $SU(N)^2  {\tilde U}(1)$   
are matched. 

In conclusion, the double c-f  locking phase,  with massless baryons 
$\hat{B}^{A, B}$, or  $\tilde{B}^{A, B}$,  or analogous states with  $1 \leftrightarrow 2$,   together with 
 $ B^{AB,\, 1} $ and  $ B^{AB,\, 2}$ (both antisymmetric in $AB$),   
 is consistent with anomaly matching.    The asymmetric way $\psi^{ij, 1}$ and $\psi^{ij, 2}$ appears in the IR baryons is consistent as the $SU(2)$ is broken. 

\subsection{Phase with unbroken $SU(2)$ \label{unbrokenSU2}}

Another phase is the one with an unbroken $SU(2)$ symmetry.
Assume (\ref{condens11})
with the same coefficients
\be
c=c' \ .
\ee
The symmetry is broken to
\be
SU(N)_{\rm cf} \times  {\tilde U}(1) \times SU(2)_{\rm ff} \times SU(8)\;,\label{thistime2}
\ee
where $SU(2)_{\rm ff} $ is a linear combination of  $SU(2)_{\rm f}$  and 
\be  SU(2) \subset SU(2N)\subset SU(2N+8)\;   \label{su2ff} 
\ee 
which iexchange the first and second $N$ flavors.  The charges of the unbroken $SU(2)$ are:
\be       \left(\begin{array}{c}\psi^{ij, 1} \\\psi^{ij, 2}\end{array}\right)\sim {\underline 2}\;,\qquad  
\left(\begin{array}{c}\eta_i^{A \le N} \\\eta_i^{N\le A\le 2N}\end{array}\right)  \sim {\underline 2^*}\;.  \label{su2fff} 
\ee
The ${\tilde U}(1)$ charges are as before,  
\be     \psi:    \quad 1  \;,\qquad \eta^{B\le 2N} :\quad -1\;,\qquad  \eta^{B> 2N} :\quad -\frac{1}{2}\;.
\ee 
The baryons are 
\be   B^{A, C}=   \sum_{i,j} ( \psi^{ij, 1}   \eta_i^{A\le N} \eta^C_j +  \psi^{ij, 2}   \eta_i^{N<A\le 2N} \eta^C_j )\;,\qquad C> 2N\;,
\ee
which is a $SU(2)$ singlet;  the others are
\bea &&  B^{[A_1 B_1],\, 1} =    \psi^{ij\;, 1}  \eta_i^{A_1}  \eta_j^{B_1}\;,   \qquad   A_1,B_1 = 1,2,\ldots, N \ , \nonumber \\ && B^{[A_2 B_2],\, 2} =    \psi^{ij\;, 2}  \eta_i^{A_2}   \eta_j^{B_2}\;,   \qquad   A_2, B_2 = N+1,N+2,\ldots, 2N  \ ,
\eea
which form a doublet.  Their ${\tilde U}(1) $ charges are:
\be     B^{A, C}: \quad   -\frac{1}{2}\;;\qquad    B^{[AB],\, m}:\quad   -1\;. 
\ee
The anomaly saturation can be again seen quickly by inspecting Table~\ref{unbrokenFFsu2}.
 \begin{table}[h!t]
  \centering 
  \begin{tabular}{|c|c|c |c|c|c|  }
\hline
 $ \phantom{{{   {  {\yng(1)}}}}}\!  \! \! \! \! \!\!\!$   & fields  &  $SU(N)_{\rm cf} $    &  $ SU(8)$    & $SU(2)$    &   $ {\tilde U}(1)   $  \\
 \hline
   \phantom{\huge i}$ \! \!\!\!\!$  {\rm UV}&  $\psi$   &   $2 \cdot   { \yng(2)} $  &    $N(N+1)\cdot (\cdot) $    &   $  \frac{N(N+1)}{2}  \,\cdot \, \yng(1)$    & $1$    \\
& $ \eta^{A_i}$      &   $  2   \cdot  ( {\bar  {\yng(2)}}  \oplus   {\bar  {\yng(1,1)}}  ) $ &    $ 2 N^2 \cdot (\cdot) $    & $N^2 \, \cdot  \, \yng(1)$     &   $ - 1$ \\
&  $ \eta^{C}$      &  $ 8 \cdot     {\bar  {\yng(1)}}  $     &   $ N \cdot  {\yng(1)}$      &    $8N \cdot (\cdot) $       &    $ - \frac{1}{2} $   \\
  \hline 
     $ \phantom{{\bar{ \bar  {\bar  {\yng(1)}}}}}\!  \! \! \! \! \! \!\!\!$   {\rm IR}&  $ B^{A,  C}$      &   $ 8 \cdot     {\bar  {\yng(1)}}   $     &   $  N \cdot    {\yng(1)}$      &    $8N \cdot (\cdot) $       &    $ - \frac{1}{2} $   \\
  &  $ B^{[A_i B_i],m}$      &  $  2   \cdot  {\bar  {\yng(1,1)}}   $      &  $ N(N-1)\cdot  (\cdot)   $     &  $\frac{N(N-1)}{2}  \,   \yng(1)$        &    $ - 1 $   \\
\hline
\end{tabular}
  \caption{\footnotesize  An $SU(2)$ flavor-flavor locked symmetric phase in the $(2,0)$  model, discussed in Subsection~\ref{unbrokenSU2}.
  $A_i$ or $B_i$  ($i=1,2$) indicate the flavor indices up to $2N$,  $C$  the rest, $2N+1, \ldots, 2N+8$.    }\label{unbrokenFFsu2}
\end{table}
 The discrete symmetries   $ {\mathbbm Z}_{\psi}=   {\mathbbm Z}_{2(N+2)}$ and  ${\mathbbm Z}_{\eta}=   {\mathbbm Z}_{2(N+4)}$ are both broken by the condensates.  
Also, Witten's  $SU(2)$ anomaly matches:   there are 
\be \frac{N(N+1)}{2}+ N^2 
\ee
left-handed $SU(2)$ doublets in the UV,  whereas the corresponding number in the IR is
\be     \frac{N(N-1)}{2}\;:
\ee
the difference is 
\be    N(N+1)\;, 
\ee
which is always even.

\subsubsection{Remarks on less symmetric phases \label{sec:less}}

The less symmetric phases  discussed  in Subsection~\ref{possible2} can be derived from the most symmetric phase 
discussed here.  Namely, when the bi-fermion condensates have no special relations,  some of the global symmetries are broken, and 
a multiplet (irrep) with respect to such a subgroup (e.g.,   $SU(N)_{\rm cf} $ or $SU(2)$) is replaced by a simple multiplicity of states of similar types, 
both for elementary fermions and for composite ones.   Clearly the anomaly saturation valid in the most symmetric case imply similar results 
by subset of fermions / subgroups  in the less symmetric phases.

 \section{ $(N_{\psi},N_{\chi}) = (3,0)  $  }
\label{cinque}

Let us consider a further generalization.  The matter fermions are
\beq
   \psi^{\{ij,\, m\}}\,, \quad    \eta_i^B\, , \qquad  m=1,2,3 \;, \quad  B=1,2,\ldots , 3(N+4)\,,
\eeq
or
\be      3\,  \yng(2) +   3(N+4) \, {\bar {\yng(1)}}\;. 
\ee
The (continuous) symmetry of this model is 
\beq
SU(N)_{\rm c} \times SU(3)_{\rm f} \times SU(3N+12)_{\rm f} \times {U}(1)\,,
    \eeq
    where $U(1)$ is the anomaly-free combination of  $U_{\psi}(1)$ and $U_{\eta}(1)$,
     \be  U(1): \qquad      \psi \to e^{i \alpha / 3 (N+2)} \psi\;, \qquad  \eta \to  e^{-   i \alpha / 3 (N+4)} \eta\;.  \label{combBis}
    \ee
   The first coefficient of the beta function is
     \beq    
 b_0= \frac 13\left[11 N   -  3  (N+ 2)  -  3  (N+4) \right] =  \frac{ 5 N  -  18}{3}  \;,
\eeq
which is positive for  $N \ge 4$. It can be seen that, as for the $(2,0)$ model, chiral symmetric phase and partial color-flavor locking do not provide solutions to the anomaly matching.

%

\subsection{Triple  color-flavor locking}

Generalizing  Subsection~\ref{possible2}, one may assume a triple color-flavor locking here: 
\bea &&  \brc  \psi^{\{ij\,, 1\}}   \eta_i^B \ckt =    \, c  \, \Lambda^3 \delta^{j \, B}\;,   \qquad  \quad  \quad j, B=1,2,\dots  N\;,   \nonumber \\
  && \brc  \psi^{\{ij\,, 2\}}   \eta_i^B \ckt =   \, c'  \, \Lambda^3  \delta^{j, \,, B-N}\;,   \qquad \,   j, B=N+1,N+2,\dots  2N\;.    
  \nonumber \\   && \brc  \psi^{\{ij\,, 3\}}   \eta_i^B \ckt =   \, c''  \, \Lambda^3 \delta^{j, \, B-2N}\;,   \qquad   j, B=2N+1,2N+2,\dots  3N\;.    \label{condens333}
\eea
The symmetry realization is 
\be  
SU(N)_{\rm cf} \times  {\tilde U}(1) \times SU(12)\;.\label{thistime3}
\ee
where ${\tilde U}(1)$ acts as 
\bea  &&  \ \,  \psi \to e^{i \tfrac{\alpha} { 3 (N+2)} } \psi\;; \nonumber \\
   && \eta^A_i \to   e^{- i \tfrac{\alpha }{ 3 (N+2)}}    \eta^A_i\;, \qquad  A=1,2, \ldots, 3N\;;
\nonumber \\&&  \eta^A_i \to   e^{  - i \tfrac{\alpha }{ 6 (N+2)} }  \eta^A_i\;, \qquad   A=3N+1,3N+2, \ldots, 3N+12\;.
\eea
or by renormalizing the charges:
\be    {\tilde Q}_{\psi} : \quad   1\;;\qquad   {\tilde Q}_{\eta^<} : \quad   -1\;;\qquad    {\tilde Q}_{\eta^>} : \quad   -\frac{1}{2}\;.
\ee

We now check the matching with massless baryons
\bea && \ \ \  \hat{B}^{A, B}=      \psi^{ij\;, 1}  \eta_i^A    \eta_j^B\,,   \qquad   A = 1,2,\ldots, N\,,    \quad   B=3N+1, \ldots 3N+12\;, \nonumber \\
 &&  B^{[AB],\, 1} =    \psi^{ij\;, 1}  \eta_i^A   \eta_j^B\;,   \qquad   A,B = 1,2,\ldots, N  \;,
 \nonumber \\&&  B^{[AB],\, 2} =    \psi^{ij\;, 1}  \eta_i^A   \eta_j^B\;,   \qquad   A,B = N+1,N+2,\ldots, 2N  \;,
 \nonumber \\  && B^{[AB],\, 3} =    \psi^{ij\;, 1}  \eta_i^A   \eta_j^B\;,   \qquad   A,B = 2N+1,2N+2,\ldots, 3N  \;.
\eea
The  ${\tilde U}(1) $ charge of these baryons are:
\be     \hat{B}^{A, B}:    \qquad   -  \frac{1 }{ 2}\;; 
\ee
\be  B^{[AB],\, 1},\quad B^{[AB],\, 2} \;, \quad  B^{[AB],\, 3} \;:     \qquad       - 1\;.
\ee
It can be readily verified that the anomalies with respect to    $SU(12)^3$, $SU(12)^2 {\tilde U}(1)$,  ${\tilde U}(1)^3$, $ {\tilde U}(1)$, $SU(N)^3$,
$SU(N)^2  {\tilde U}(1)$  agree in the UV and in the IR.

\subsection{$SU(3)$ symmetric phase  \label{su3symmetric}}

As in Subsection \ref{unbrokenSU2} one may assume a more symmetric form of the condensates (\ref{condens333})
with equal coefficient
\be
c=c'=c'' \ .
\ee
  In this case,  a diagonal $SU(3)$ between $SU(3)_{\psi}$ and  $SU(3) \subset  SU(3N)$ remains unbroken.
The  low-energy symmetry realization is then 
\bqa  
 SU(N)_{\rm cf} \times  SU(3) \times   {\tilde U}(1) \times SU(12)\;.\label{thistime4}
\eea
There are two more triangles,   $SU(3)^3$ and $SU(3)^2 {\tilde U}(1)$, in addition to six types of anomalies considered in the previous Subsection.  
The charges with respect to this $SU(3)$ are
\bqa       \left(\begin{array}{c}\psi^{ij, 1} \\\psi^{ij, 2}  \\ \psi^{ij, 3}   \end{array}\right)     \sim {\underline 3}\;,\qquad  
\left(\begin{array}{c}   \eta_i^{A \le N} \\ \eta_i^{N<A\le 2N}  \\ \eta_i^{2N < A \le 3N}   \end{array}\right)       \sim {\underline 3^*}\;.
\eea
The massless baryons are
\be   B^{A, C}=   \sum_{i,j} ( \psi^{ij, 1}   \eta_i^{A\le N} \eta^C_j +  \psi^{ij, 2}   \eta_i^{N<A\le 2N} \eta^C_j  +  \psi^{ij, 3}   \eta_i^{2N<A\le 3N} \eta^C_j )    \;,\qquad C> 3N\;,
\ee
which is an $SU(3)$ singlet;  the others are
\bea  && B^{[AB],\, 1} =    \psi^{ij\;, 1}  \eta_i^A   \eta_j^B\;,   \qquad   A,B = 1,2,\ldots, N  \ ,
 \nonumber \\ &&  B^{[AB],\, 2} =    \psi^{ij\;, 2}  \eta_i^A   \eta_j^B\;,   \qquad   A,B = N+1,N+2,\ldots, 2N  \ ,
 \nonumber \\  && B^{[AB],\, 3} =    \psi^{ij\;, 2}  \eta_i^A   \eta_j^B\;,   \qquad   A,B = 2N+1,2N+2,\ldots, 3N  \ ,
\eea
which form an anti-triplet, ${\underline 3^*}$.

Again it is convenient to have the decomposition of the fields with respect to the unbroken groups.  The saturation of the anomalies 
 $SU(12)^3$,  $SU(12)^2 {\tilde U}(1)$, ${\tilde U}(1)^3$, ${\tilde U}(1)$, $SU(N)^3$, $SU(N)^2  {\tilde U}(1)$, $SU(3)^3$, and   $SU(3)^2  {\tilde U}(1)$ 
 is seen at once, by inspection of  Table~\ref{decompositionbis}. 
 \begin{table}[h!t]
  \centering 
  \begin{tabular}{|c|c|c |c|c|c|  }
\hline
$ \phantom{{{   {  {\yng(1)}}}}}\!  \! \! \! \! \!\!\!$   & fields  &  $SU(N)_{\rm cf} $    &  $ SU(12)$    & $SU(3)$    &   $ {\tilde U}(1)   $  \\
 \hline
 \phantom{\huge i}$ \! \!\!\!\!$  {\rm UV} &  $\psi$   &   $3 \cdot   { \yng(2)} $  &    $   \frac{3  N(N+1)}{2}  \,\cdot   (\cdot) $    &   $  \frac{N(N+1)}{2}  \,\cdot \,  {\yng(1)}$    & $ 1 $    \\
& $ \eta^{A\le 3N}$      &   $  3   \cdot  ( {\bar  {\yng(2)}}  \oplus   {\bar  {\yng(1,1)}}  ) $ &    $3 N^2 \cdot (\cdot) $    & $N^2 \, \cdot  \,{\bar  {\yng(1)}}$     &   $ -1 $ \\
&  $ \eta^{A>3N}$      &  $ 12 \cdot     {\bar  {\yng(1)}}  $     &   $ N \cdot  {\yng(1)}$      &    $   12 N \cdot  (\cdot) $       &    $ - \frac{1}{2} $   \\
  \hline 
  $ \phantom{{\bar{ \bar  {\bar  {\yng(1)}}}}}\!  \! \! \! \! \! \!\!\!$   {\rm  IR}&   $ B^{A_<,C_>}$      &   $ 12  \cdot     {\bar  {\yng(1)}}   $     &   $  N \cdot    {\yng(1)}$      &    $ 12 N \cdot  (\cdot)  $       &    $ - \frac{1}{2} $   \\
 &   $ B^{[AB],m}$      &  $  3   \cdot  {\bar  {\yng(1,1)}}   $    &  $ \frac{3 N(N-1)}{2}   \cdot  (\cdot)  $     &  $\frac{N(N-1)}{2}  \,    {\yng(1)}$        &    $ - 1 $   \\
\hline
\end{tabular}  
  \caption{\footnotesize  Color-favor-flavor locked $SU(3)$ symmetric phase in the $(3,0)$ model, discussed in Subsection~\ref{su3symmetric} 
}\label{decompositionbis}
\end{table}

As seen in the $(2,0)$ model, Subsection~\ref{sec:less},    less symmetric phases are possible in the $(3,0)$ model as well.  The condensates (\ref{condens333})
are of more general forms in those cases, with unequal values, and  one or both of the symmetries $SU(N)_{\rm cf} $ and  $SU(3)$    can be broken spontaneously. 
The set of the baryons $ B^{A,C}$  and  $ B^{[AB],m}$  will continue to saturate the anomaly triangles of the remaining symmetries.

\section{$(N_{\psi},N_{\chi}) = (0,1)$ }
\label{sei}

Let us  review the $(N_{\psi},N_{\chi}) = (0,1)$ model studied in \cite{Dimopoulos:1980hn,ACSS,ADS,Poppitz,ShiShr,BKS}. The matter fermions are 
 \beq
   \qquad   \chi_{[ij]}  \;, \qquad {\tilde  \eta}^{B\, j} \;,  \qquad   \quad  B=1,2,\ldots, (N-4) \,.
\eeq
The first coefficient of the $\beta$ function is
\be  b_0=  \frac 13\left[11 N -  (N-2) -  (N-4) \right]= \frac{ 9N+6}{3}\;. 
\ee
The  (continuous) symmetry is
\be     SU(N)_{\rm c}  \times  SU(N-4)_{\rm f} \times U(1)\;,
\ee
where  the anomaly free $U(1)$ charge  is 
\be   \chi:    \quad    N-4  \;; \qquad {\tilde  \eta}^{B\, j} \;:  \quad   - (N-2)  \;. 
\ee
There are also discrete symmetries 
    \be    {\mathbbm Z}_{\chi}  =  {\mathbbm Z}_{N-2}  \subset  U_{\psi}(1)\;, \qquad    {\mathbbm Z}_{\eta}  ={\mathbbm Z}_{N-4}  \subset  U_{\eta}(1)\;. 
    \ee

\subsection{Chirally symmetric phase   in the $(0,1)$ model}

    Let us first examine the possibility that no condensates form, the system confines and the flavor symmetry is unbroken \cite{Dimopoulos:1980hn}.
 The massless baryons are   
 \be      B^{\{CD\}} = \chi_{[ij]} \,  \tilde{\eta}^{i\, C}   \tilde{\eta}^{j\, D}   \;, \qquad   C,D=1,2,\ldots (N-4)\;,  \label{massless}\ee
 symmetric in $C \leftrightarrow D$.   They have the ${U}(1)$ charge: $ -N$.
 The matching of the anomalies can be read off Table~\ref{01model}.
\begin{table}[h!t]
  \centering 
  \begin{tabular}{|c|c|c |c|c|  }
\hline
 $ \phantom{{{   {  {\yng(1)}}}}}\!  \! \! \! \! \!\!\!$   & fields   &  $SU(N)_{\rm c}  $    &  $ SU(N-4)$     &   $ U(1)   $  \\
 \hline 
  $ \phantom{{\bar{ \bar  {\bar  {\yng(1,1)}}}}}\!  \! \!\! \! \!  \!\!\!$  {\rm UV}&  $\chi$   &   ${\bar  { \yng(1,1)}}   $  &    $  \frac{N(N-1)}{2} \cdot (\cdot) $    & $N-4$    \\
& $ {\tilde \eta}^{A}$      &   $  (N-4)  \cdot   { {\yng(1)}}   $     & $N \, \cdot  \, {\yng (1)}  $     &   $ - (N-2)  $ \\
   \hline     
  \phantom{\huge i}$ \! \!\!\!\!$  {\rm IR}&    $ B^{\{AB\}}$      &  $  \frac{(N-4)(N-3)}{2} \cdot ( \cdot )    $         &  $ {\yng(2)}$        &    $ - N $   \\
\hline
\end{tabular}
  \caption{\footnotesize  Confinement and unbroken symmetry in the $(0,1)$ model} 
  \label{01model}
\end{table}

\subsection{Color-flavor locked vacuum  \label{remark01}}

It was pointed out \cite{ADS} that this system may instead develop a condensate of the form
\be   \brc   \chi_{[ij]} {\tilde  \eta}^{B\, j}   \ckt  = \const   \, \Lambda^3 \delta_i^B \;, \qquad i, B=1,2,\ldots, N-4\;,  \label{cfl01}
\ee
namely,
\be   {\bar  {\yng(1,1)} }  \otimes  \yng (1)  \to    {\bar  {\yng(1)} }  \oplus  \ldots \;.
\ee
The symmetry is broken to 
\beq
   SU(N-4)_{\rm cf}  \times  U^{'}(1)   \times   SU(4)_{\rm c}\,.
    \label{b270}
\eeq
The massless baryons (\ref{massless})  saturate all the anomalies associated with $SU(N-4)_{\rm cf}  \times  U^{'}(1)$.
There remains the  $\chi_{i_2 j_2}$  fermions which remain massless and strongly coupled to the $SU(4)_{\rm c}$.  We may assume that 
$SU(4)_{\rm c}$ confines, and the condensate
\be  \brc  \chi \chi \ckt \ne 0\;, 
\ee
in
 \be    {\bar {  \yng(1,1) }} \otimes  {\bar {  \yng(1,1) }}  \, \to \,  {\bar { \yng(1,1,1,1) }}  \, \oplus  \ldots \;,
\ee
forms and $\chi_{i_2 j_2}$ acquire dynamically mass. 
Assume that the massless baryons are:
\be      B^{\{A B\}} = \chi_{[ij]} \,  \tilde{\eta}^{i\, A}   \tilde{\eta}^{j\, B}   \;, \qquad   A,B=1,2,\ldots (N-4)\;, 
\ee
 the saturation of all of the triangles associated can be seen in Table~\ref{SimplestAgain2}.  
\begin{table}[h!t]
  \centering 
  \begin{tabular}{|c|c|c |c|c|  }
\hline
$ \phantom{{{   {  {\yng(1)}}}}}\!  \! \! \! \! \!\!\!$   & fields     &  $ SU(N-4)_{\rm cf} $     &   $ U^{\prime}(1) $     &  $SU(4)_{\rm c}  $     \\
 \hline
   $ \phantom{{\bar{ \bar  {\bar  {\yng(1,1)}}}}}\!  \! \!\! \! \!  \!\!\!$  {\rm UV}&  $\chi_{i_1 j_1}$     &    $  {\bar  { \yng(1,1)}}   $    & $N$   &   $\frac{(N-4)(N-5)}{2}\cdot (\cdot)  $ \\
 &  $\chi_{i_1 j_2}$   &    $  4   \cdot {\bar  { \yng(1)}} $    & $\frac{N}{2}$   &   $ (N-4) \cdot {\bar  { \yng(1)}}   $     \\
 &$\chi_{i_2 j_2}$   &    $  \frac{4 \cdot 3}{2} \cdot (\cdot) $    & $0$    &   ${\bar  { \yng(1,1)}}   $     \\
& $ {\tilde \eta}^{A, \,i_1}$          & $\yng(2) \oplus \yng(1,1)$     &   $ - N $    &   $  (N-4)^2  \cdot  (\cdot)   $  \\
 & $ {\tilde \eta}^{A, \,i_2}$         & $4\, \cdot  \, {\yng (1)}  $     &   $ - \frac{N}{2} $     &   $  (N-4)  \cdot  \yng(1)  $  \\
   \hline 
     \phantom{\huge i}$ \! \!\!\!\!$  {\rm IR}&     $ B^{\{AB\}}$        &  $ {\yng(2)}$        &    $ - N $     &  $  \frac{(N-4)(N-3)}{2} \cdot ( \cdot )    $      \\
\hline
\end{tabular}
  \caption{\footnotesize  Color-flavor locking  in the $(0,1)$ model.    The color index $i_1$ or $j_1$  runs up to $N-4$ and the rest is indicated by $i_2$ or $j_2$.}\label{SimplestAgain2}
\end{table}
 The complementarity \cite{Fradkin,BKS} does work here.

 \section{ $(N_{\psi},N_{\chi}) = (0,2)$  }
\label{sette}

 Let us now consider a generalization of the $\chi \tilde {\eta}$ model with
 \beq
   \qquad   \chi_{[ij]}^m  \;, \qquad {\tilde  \eta}^{B\, j} \;,  \qquad    m=1,2,\,\quad  B=1,2,\ldots, 2(N-4) \,.
\eeq
or
\be        2\, {\bar {\yng(1,1)}} +   2(N-4) \, { {\yng(1)}}\;. 
\ee
The first coefficient of the $\beta$ function is
\beq
  b_0 =   \frac 13\left[11 N   -  2 (N-2)  -  2 (N-4)\right] =    \frac 13  (7N +12)\;.
\eeq

\subsection{ No chiral symmetry breaking in the $(0,2)$ model?}

The symmetry is
\be   SU(N)_{\rm c}  \times    SU(2)_{\rm f} \times  SU(2N-8)_{\rm f} \times U(1)\;,
\ee
where  the anomaly free $U(1)$ charge  is 
\be   \chi:    \quad   {N-4} \;; \qquad {\tilde  \eta}^{B\, j} \;:  \quad   -(N-2)\;.   \label{anfru10}
\ee
Let us assume that the massless baryons are   
 \be      B^{\{CD\},m} = \chi_{[ij]}^m \,  \tilde{\eta}^{i\, C}   \tilde{\eta}^{j\, D}  \;   \;, \qquad   C,D=1,2,\ldots 2(N-4)\;, \quad m=1,2\;. \ee
 symmetric in $CD$.    They have the ${U}(1)$ charge $  -N $.
There is no way  $  B^{\{CD\}} $ can saturate the anomalies in  $SU(2N-8)_{\rm f} \times U(1)$.

One concludes that the confinement phase with unbroken chiral symmetry 
$SU(2)\times SU(2N-8)_{\rm f} \times U(1)$  is not possible. This is, again,  in contrast to the $(N_{\psi},N_{\chi}) = (0,1)$ model.

\subsection{ Color-flavor locking}

Let us instead   assume a color-flavor locked  diagonal  VEV,
\bea &&
   \langle   \chi_{[ij]}^1 \,  \tilde{\eta}^{i\, B}  \rangle =      \, c \,  \Lambda^3   \delta_j^B \,, \qquad  \qquad \ j, \, B =1,2, \ldots ,  N-4\;, \nonumber \\
  && \langle   \chi_{[ij]}^2 \,  \tilde{\eta}^{i\, B}  \rangle =      \, c'  \,  \Lambda^3   \delta_j^{B- (N-4)} \,,  \ \quad    j =1, \ldots ,  N-4\;;\quad 
  \, B =N-3\ldots ,  2N-8\;.   
 \label{cond2222}
\eea
Then the symmetry is broken to
\beq
  SU(N-4)_{\rm cf}  \times U^{'}(1) \times   SU(4)_{\rm c}\,.
    \label{b27}
\eeq
where   $U(1)^{'} $ is the unbroken  linear combination between 
   the anomaly free $U(1)$, (\ref{anfru10}), and a subgroup of the color $SU(N)$,  ${\rm diag} (   \tfrac{1}{N-4} {\mathbbm 1}_{N-4}, -     \tfrac{1}{4} {\mathbbm 1}_{4})$. 
   We assume that the massless baryons are
    \bea  &&    B^{\{CD\},1} = \chi_{[ij]}^1 \,  \tilde{\eta}^{i\, C}   \tilde{\eta}^{j\, D}  \;   \;, \qquad   C,D=1,2,\ldots N-4\;, \nonumber \\    &&  {\hat B}^{\{CD\},1} = \chi_{[ij]}^1 \,  \tilde{\eta}^{i\, C}   \tilde{\eta}^{j\, D}  \;   \;, \qquad   C,D=N-3,N-2,\ldots 2(N-4)\;.\eea
   The charges under    $  SU(N-4)_{\rm cf} \otimes U^{'}(1)$
are given in Table \ref{SimplestAgain02} where the  $U(1)^{'}$ charges are appropriately renormalized by a common factor.   
All anomalies  $SU(N-4)_{\rm cf}^3$,  ${U(1)^{'}}^3$,  $U(1)^{'}$, $U(1)^{'}  SU(N-4)_{\rm cf}^2$  work out fine. 
\begin{table}[h!t]
  \centering 
  \begin{tabular}{|c|c|c |c|  }
\hline
$ \phantom{{{   {  {\yng(1)}}}}}\!  \! \! \! \! \!\!\!$   & fields     &  $ SU(N-4)_{\rm cf} $     &   $ U^{\prime}(1) $      \\
 \hline
   $ \phantom{{\bar{ \bar  {\bar  {\yng(1,1)}}}}}\!  \! \!\! \! \!  \!\!\!$  {\rm UV}&  $\chi^1_{[i_1 j_1]}$     &    $  {\bar  { \yng(1,1)}}   $    & $1$  \\
 &  $\chi^1_{[i_1 j_2]}$   &    $  4   \cdot {\bar  { \yng(1)}} $    & $\frac{1}{2}$   \\
 &$\chi^1_{[i_2 j_2]}$   &    $  (\cdot) $    & $0$    \\
& $ {\tilde \eta}^{B, \,i_1}$          & $ 4(N-4)  \,\,   {\yng(1)}$     &   $ - 1 $    \\
 & $ {\tilde \eta}^{B, \,i_2}$         & $8\, \cdot  \, {\yng (1)}  $     &   $ - \frac{1}{2} $      \\
   \hline 
     \phantom{\huge i}$ \! \!\!\!\!$  {\rm IR}&    $ B^{\{CD\, 1\}}    $     &  $ {\yng(2)}$        &    $ - 1 $      \\&     $  {\hat B}^{\{CD\, 1\}}     $     &  $ {\yng(2)}$        &    $ - 1 $         \\
\hline
\end{tabular}
  \caption{\footnotesize  Color-flavor locking  in the $(0,2)$ model.    The color index $i_1$ or $j_1$  runs up to $N-4$. 
  The rest is indicated by $i_2$ or $j_2$.}\label{SimplestAgain02}
\end{table}

\subsection{ Phase with unbroken $SU(2)$ }

Assume instead  that the condensates (\ref{cond2222}) occur with 
with the same coefficients
\be
c=c' \ .
\ee
Then the residual symmetry is bigger
\beq
SU(N-4)_{\rm cf}  \times SU(2) \times  U^{'}(1) \times   SU(4)_{\rm c}\,.
    \label{b272}
\eeq

The baryons are
\bea     &&  B^{\{CD\},1} = \chi_{[ij]}^1 \,  \tilde{\eta}^{i\, C}   \tilde{\eta}^{j\, D}  \;   \;, \qquad   C,D=1,2,\ldots N-4\;, \nonumber \\ &&     {B}^{\{CD\},2} = \chi_{[ij]}^2 \,  \tilde{\eta}^{i\, C}   \tilde{\eta}^{j\, D}  \;   \;, \qquad   C,D=N-3,N-2,\ldots 2(N-4)\;,\eea
symmetric in $CD$.
The charges with respect to this $SU(2)$ are
\bqa       \left(\begin{array}{c}\chi_{[ij]}^{1} \\\chi_{[ij]}^{2}   \end{array}\right)     \sim {\underline 2}\;,\qquad  
\left(\begin{array}{c}   {\tilde \eta}_i^{A \le N-4} \\ {\tilde \eta}_i^{N-4 <A\le 2N-8}     \end{array}\right)       \sim {\underline 2^*}\;.
\eea
The charges of the fields with respect to  the unbroken symmetries are in Table~\ref{decompositionBis}. The saturation of all seven  types of triangles  can be seen by inspection.
{\tiny
 \begin{table}[h!t]
  \centering 
  \begin{tabular}{|c|c|c |c|c|c|  }
\hline
$ \phantom{{{   {  {\yng(1)}}}}}\!  \! \! \! \! \!\!\!$   & fields   &  $SU(N-4)_{\rm cf} $    &  $ SU(2)$    &  $ U^{'}(1)   $   &  $SU(4)_{\rm c}$   \\
 \hline
 $ \phantom{{\bar{ \bar  {\bar  {\yng(1,1)}}}}}\!  \! \!\! \! \!  \!\!\!$   {\rm UV}&  $\chi^m_{[i_1 j_1]} $   &   $ 2 \cdot {\bar  { \yng(1,1)}} $  &    $   \frac{(N-4)(N-5)}{2} \cdot  \yng(1) $    &   $ 1$    & $  (N-4)(N-5)  \,\cdot \, (\cdot) $    \\
 &$   \chi^m_{[i_1 j_2]} $      &   $  8   \cdot   {\bar  {\yng(1)}}     $ &    $  4   (N-4)\cdot \yng(1)$    & $ {\frac{1}{2}}$     &   $ 2   (N-4) \cdot  {\bar  {\yng(1)}}   $ \\
 &  $      \chi^m_{[i_2 j_2]} $      &   $  12  \cdot  (   \cdot  ) $ &    $ 6 \cdot  \yng(1) $    & $    0 $     &   $ 2 \cdot  {\bar  {\yng(1,1)}}    $ \\
 & $  {\tilde \eta}^{B\, i_1}      $      &  $ 2 \cdot    (  {\yng(1,1)}+    {\yng(2)} ) $     &   $  (N-4)^2  \cdot  {\bar {\yng (1)}}$      &    $-1 $       &    $ 
  2(N-4)^2  \cdot  (\cdot)  $   \\
  & $  {\tilde \eta}^{B\, i_2}    $      &  $ 8    \cdot     {{\yng(1)}}  $     &   $ 4(N-4)  \cdot  {\bar {\yng (1)}}$      &    $- \frac{1}{2} $       &    $2 (N-4)  \cdot \yng (1)$   \\
  \hline 
   $ \phantom{{\bar{ \bar  {\bar  {\yng(1)}}}}}\!  \! \! \! \! \! \!\!\!$    {\rm IR}&  $ B^{CD, m}$      &   $ 2   \cdot     {{\yng(2)}}   $     &   $  \frac{(N-4)(N-3)}{2}  \cdot    {\bar {\yng (1)}}$      &    $    -1$       &    $ (N-4)(N-3)  \cdot 
   ( \cdot ) $   \\
\hline
\end{tabular}
    \caption{\footnotesize $SU(2)$ symmetric phase in the $(0,2)$ model.   
  $i_1,  j_1$  stand for  the color indices up to  $N-4$,    $i_2, j_2$ the last four.    }\label{decompositionBis}
\end{table}
 }

$SU(2)$ has no (perturbative) triangle anomaly  but it does have a global  anomaly (Witten).  It can be readily checked that the difference of the number 
of the doublets in the UV and in the IR is even.

As in the $(0,1)$ model,  the fermions  $ \chi^m_{[i_2j_2]} $   remain massless and coupled strongly by the unbroken color 
 $SU(4)_{\rm c}$.   It is possible that they condense  as
  \be    \brc  \epsilon^{ijk\ell}  \chi_{ij}^m  \chi_{k \ell}^n  \ckt  \ne 0  \;, \qquad m,n =1,2.
 \ee
   As they are symmetric in $m,n$,    the symmetry is broken as
 \be   SU(2)\to SO(2)=U(1)\;, 
 \ee
 in a scenario similar to tumbling. 
 
 So after all   $SU(2)$ is dynamically broken.  The fate of the unbroken, residual  $SU(4)_{\rm c}$ is similar to what happens in the second (XSB) scenario 
 in Subsection \ref{remark01}.

 \section{ $(N_{\psi},N_{\chi}) = (0,3)$  }
\label{otto}

 The model to be considered now is 
\beq
   \qquad   \chi_{[ij]}^m  \;, \qquad {\tilde  \eta}^{B\, j} \;,  \qquad    m=1,2,3,\,\quad  B=1,2,\ldots, 3(N-4) \,.
\eeq
or
\be        3 \, {\bar {\yng(2)}} +   3 (N-4) \, { {\yng(1)}}\;. 
\ee
The first coefficient of the $\beta$ function is
\beq
  b_0=   \frac 13\left[11 N   -  3 (N-2)  -  3 (N-4)\right] =    \frac 13  (5 N +18)\;.
\eeq 
The symmetry is
\be    SU(N)_{\rm c}  \times   SU(3)\times  SU(3N-12)_{\rm f} \times U(1)\;,
\ee
where  the anomaly free $U(1)$ charge  is 
\be   \chi:    \quad    N-4   \;, \qquad {\tilde  \eta}^{B\, j} \;:  \quad   -  (N-2)   \;.   \label{anfru1}
\ee
Again, the option of confinement with no flavor symmetry breaking is excluded. 

\subsection{Color-flavor locking  \label{CF03}}  

Let us try to generalize the color-flavor locking of the  $(N_{\psi},N_{\chi}) = (0,2)$ case to our $(N_{\psi},N_{\chi}) = (0,3)$  model,   by assuming
 \bea 
   && \langle   \chi_{[ij]}^1 \,  \tilde{\eta}^{i\, B}  \rangle =      \, c \,  \Lambda^3   \delta_j^B \neq 0\,, \qquad \qquad \, \    j, \, B =1,2, \ldots ,  N-4\;, \nonumber \\   
 \label{cond333}
&&  \langle   \chi_{[ij]}^2 \,  \tilde{\eta}^{i\, B}  \rangle =      \, c  \,  \Lambda^3   \delta_j^{B- (N-4)} \neq 0\,, \qquad  
  j =1,2, \ldots ,  N-4\;,\quad 
  \, N-3\le  B \le   2N-8\;,
 \label{cond444}
\nonumber \\   
     && \langle   \chi_{[ij]}^3 \,  \tilde{\eta}^{i\, B}  \rangle =      \, c
        \,  \Lambda^3   \delta_j^{B- (N-4)} \neq 0\,, \qquad  
  j =1,2, \ldots ,  N-4\;,\quad 
  \, 2N-7 \le B \le 3N-12\;.   \nonumber \\  
 \label{cond555}
\eea
Then the  symmetry breaking pattern  is
\beq
 SU(N-4)_{\rm cf}  \times U^{'}(1) \times  SU(3) \times   SU(4)_{\rm c}\,.
    \label{b273}
\eeq
where   $U(1)^{'} $ is the unbroken  linear combination between 
   the anomaly free $U(1)$, (\ref{anfru1}), and a subgroup of the color $SU(N)$,  ${\rm diag} (  4\,  {\mathbbm 1}_{N-4}, -    (N-4)  {\mathbbm 1}_{4})$. 
   The would-be $SU(3)$ multiplets are:
   \bqa       \left(\begin{array}{c}\chi_{[ij]}^{1} \\\chi_{[ij]}^{2}  \\ \chi_{[ij]}^3     \end{array}\right)     \sim {\underline 3}\;,\qquad  
\left(\begin{array}{c}   {\tilde \eta}_i^{A \le N-4} \\ \eta_i^{N-4 <A\le 2N-8}  \\ \eta_i^{ 2N-8 <A\le 3N-12}   \end{array}\right)       \sim {\underline 3^*}\;.
\eea

   We assume that the massless baryons are
    \bea    &&  B^{\{CD\},1} = \chi_{[ij]}^1 \,  \tilde{\eta}^{i\, C}   \tilde{\eta}^{j\, D}  \;   \;, \qquad   C,D=1,2,\ldots,  N-4\;,  \nonumber \\   &&  {\hat B}^{\{CD\},2} = \chi_{[ij]}^2 \,  \tilde{\eta}^{i\, C}   \tilde{\eta}^{j\, D}  \;   \;, \qquad   C,D=N-3,\ldots, 2(N-4)\;,\nonumber \\   &&  {\tilde B}^{\{CD\},3} = \chi_{[ij]}^3 \,  \tilde{\eta}^{i\, C}   \tilde{\eta}^{j\, D}  \;  \;, \qquad   C,D=2N-7,\ldots, 3(N-4)\;,\eea
symmetric in $CD$.  These baryons transform as  ${\underline 3^*}$.
    \begin{table}[h!t]
  \centering 
  \begin{tabular}{|c|c|c |c|c|c|  }
\hline
$ \phantom{{{   {  {\yng(1)}}}}}\!  \! \! \! \! \!\!\!$   & fields  &  $SU(N-4)_{\rm cf} $    &  $ SU(3)$    &  $ U^{'}(1)   $   &  $SU(4)_{\rm c}$   \\
 \hline
  $ \phantom{{\bar{ \bar  {\bar  {\yng(1,1)}}}}}\!  \! \!\! \! \!  \!\!\!$  {\rm UV}&  $\chi^m_{[i_1 j_1]} $   &   $ 3 \cdot   {\bar { \yng(1,1)}} $  &    $   \frac{(N-4)(N-5)}{2} \cdot  \yng(1) $    &   $  1$    & $  \frac{3 (N-4)(N-5)}{2}   \,\cdot  (\cdot) $    \\
 &$   \chi^m_{[i_1 j_2]}  $      &   $  12   \cdot  ( {\bar  {\yng(1)}}    ) $ &    $  4   (N-4)\cdot   \yng(1)$    & ${\frac{1}{2}}$     &   $ 3   (N-4) \cdot  {\bar  {\yng(1)}}   $ \\
   &$      \chi^m_{[i_2 j_2]}  $      &   $  18  \cdot  (   \cdot  ) $ &    $ 6 \cdot  \yng(1)  $    & $  0$     &   $ 3 \cdot  {\bar  {\yng(1,1)}}    $ \\
 & $  {\tilde \eta}^{B\, i_1}   $      &  $ 3 \cdot    (  {\yng(1,1)}+    {\yng(2)} ) $     &   $  (N-4)^2  \cdot  {\bar  {\yng(1)}}$      &    $  -1 $       &    $ 3 (N-4)^2  \cdot (\cdot)  $   \\
  & $  {\tilde \eta}^{B\, i_2}    $      &  $ 12    \cdot     {{\yng(1)}}  $     &   $ 4(N-4)  \cdot  {\bar { \yng(1)}} $      &    $- \frac{1}{2} $       &    $3 (N-4)  \cdot \yng (1)$   \\
  \hline 
  $ \phantom{{\bar{ \bar  {\bar  {\yng(1)}}}}}\!  \! \! \! \! \! \!\!\!$   {\rm IR}&    $ B^{\{CD\}, m}$      &   $ 3   \cdot     {{\yng(2)}}   $     &   $  \frac{(N-4)(N-3)}{2}  \cdot    {\bar { \yng(1)}}$      &    $  -1$       &    $ \frac{3(N-4)(N-3)}{2}     \cdot ( \cdot) $   \\
\hline
\end{tabular}  
  \caption{\footnotesize  The decomposition of the fields in the $(0,3)$  model.  The color indices are divided into two groups: $i_1, j_1$ run up to $N-4$;  $i_2, j_2$ the rest. 
 Moreover, the color and flavor indices are combined as in Subsection~\ref{CF03}   }\label{decomposition3}
\end{table}

Unlike what happens to the  $(0,2)$ model, or to the $(3,0)$ model,    however,   here the unbroken $SU(3)$  symmetry cannot be realized manifestly 
in the infrared:  $SU(3)^3$  triangles do not match in the UV and  IR, see Table~\ref{decomposition3}. 

A possibility is that the
 condensates (\ref{cond333})  take unequal values. 
 With  $SU(3)$ broken,   the baryons  $ B^{\{CD\}, m}$  saturate the anomalies in
 $SU(N-4)_{\rm cf}  \times U^{'}(1)  \times   SU(4)_{\rm c}$.
 
 Another possibility is suggested by the presence of massless fermions  $ \chi^m_{[ij]}  \quad (i_>, j_>) $, which interact strongly  with the remaining 
 gauge group $SU(4)_{\rm c}$.   
It is possible that the condensates
 \be    \brc  \epsilon^{ijk\ell}  \chi_{ij}^m  \chi_{k \ell}^n  \ckt  \ne 0  \;, \qquad m,n =1,2,3.
 \ee
 form.  As they are symmetric in $m,n$,    the symmetry is broken as
 \be   SU(3)\to SO(3)\;  
 \ee
 which is free of anomalies.

\section{ $(N_{\psi},N_{\chi}) = (2,1)  $  }
\label{nove}

Next consider the $SU(N)$ gauge model with the chiral fermion sector
\beq
   \psi^{\{ij\},\, m}\;, \qquad \chi_{[ij]} \;, \qquad  \eta^B_j\;,  \qquad    m=1,2,\,\quad  B=1,2,\ldots, N + 12\,,
\eeq
or
\be      2\,  \yng(2) +  {\bar {\yng(1,1)}} +   (N+12) \, {\bar {\yng(1)}}\;. 
\ee
The symmetries of the theory are
\beq
SU(N)_{\rm c}  \times SU(2)_{{\rm f}} \times SU(12+N)_{{\rm f}} \times U(1)^2\;. 
\eeq
The two $U(1)$'s are anomaly-free combinations of $U_{\psi}(1)$, $U_{\chi}(1)$, $U_{\eta}(1)$, 
which  can be taken as 
\bea &&   U_1(1): \qquad   \psi \to  e^{i  \frac{\alpha}{2(N+2)}} \psi\;, \qquad \eta \to   e^{ -i \frac{\alpha}{N+12}} \eta\;;   
\nonumber \\
&&   U_2(1): \qquad   \psi\to  e^{i  \frac{\beta}{2(N+2)}} \psi\;, \qquad \chi \to   e^{- i \frac{\beta}{N-2}} \chi\;. \label{nonanU12}
\eea
The first coefficient of the $\beta$ function is
\beq  
b_0= \frac 13\left[ 11 \,N - 2 (N+2)-  (N-2)  -  (12+N) \right] = \frac 13\left(  7N -14 \right)\,.
\eeq

\subsection{Color-flavor locking? }
 A possibility is that a  (partial) color-flavor locking condensate
\be   \langle  \psi^{\{ij\},\, 1} \eta_{j}^B \rangle   = \, c \,  \Lambda^3 \delta^{i B} \;,   \qquad  i, B=1,2,\ldots, N  \label{no1}
\ee
 develops, where the direction of the $SU_{\psi}(2)$ breaking is  arbitrarily.  Let us assume that there is no adjoint condensate $ \langle  \psi \chi \rangle $.
 The unbroken symmetry is 
\be  
 SU(N)_{\rm cf} \times   SU(12)_{\rm f}  \times   {\tilde U}(1)\,, \label{no2}
\ee
where  $ {\tilde U}(1)$ charges are
\be    Q_{\psi}=  1\;,   \qquad  Q_{\chi}=  - \frac{N-8}{N-2}  \;, \qquad    Q_{\eta}=  -1 \ .
\ee
 The candidate baryons are:
 \be   B^{CD, m}=     \psi^{\{ij\},m}  \eta_{i}^C \eta_{j}^D \ .
 \ee
 An inspection shows that these baryons do not saturate the $G_{\rm f}$ anomalies, and one concludes that the phase (\ref{no1})
 is not possible.

 \subsection{Color-flavor-flavor locking?}
 
 Let us assume, for $N \le 12$,   the condensates of the form,
 \bea   &&  \langle  \psi^{\{ij\},\, 1} \eta_{j}^{B_1} \rangle   =  \, c \,  \Lambda^3 \delta^{i,  B_1} \;, 
  \nonumber \\  &&  \langle  \psi^{\{ij\},\, 2} \eta_{j}^{B_2} \rangle   = \, c \,  \Lambda^3 \delta^{i, B_2-N} \;, 
 \eea
 where the flavor indices $B_1$ runs up to $N$, $B_2$ from $N+1$ to $2N$.    
 The symmetry is broken to 
 \beq
SU(N)_{\rm cf}  \times SU(2)_{{\rm f f}} \times SU(12-N)_{{\rm f}} \times U^{\prime}(1)\;. 
\eeq

  \begin{table}[h!t]
  \centering 
  \begin{tabular}{|c|c|c |c|c|c|  }
\hline
$ \phantom{{{   {  {\yng(1)}}}}}\!  \! \! \! \! \!\!\!$   & fields   &  $SU(N)_{\rm cf} $    &  $ SU(2)$    &  $SU(12-N)$   &  $ U^{'}(1)   $  \\
 \hline
   \phantom{\huge i}$ \! \!\!\!\!$  {\rm UV}&  $\psi^{\{i  j\}, m} $   &   $ 2 \cdot { { \yng(2)}} $   &    $   \frac{N(N+1)}{2} \cdot  \yng(1) $    &   $ N(N+1) \cdot (\cdot)$    & $  1 $  \\
  & $\chi_{[i  j]} $   &   $  {\bar  { \yng(1,1)}} $  &    $   \frac{N(N-1)}{2} \cdot  (\cdot) $    &   $  \frac{N(N-1)}{2} \cdot  (\cdot) $    & $  -\frac{N-8}{N-2}$    \\
 &  $  {\eta}^{B_1}_i ,  {\eta}^{B_2}_i       $      &  $ 2\cdot (   {\bar  {\yng(1,1)}}\oplus   {\bar  {\yng(2)} } ) $     &   $  N^2  \cdot  {\bar {\yng (1)}}$      &    $  2  N^2 \cdot (\cdot) $       &    $ 
  -1 $   \\
 &  $  {\eta}^{B_3}_i    $      &  $ (12-N)    \cdot     {\bar {\yng(1)}}  $     &   $ N(12-N) \cdot  (\cdot)$      &    $N\cdot  {\yng(1)} $       &    $-1$   \\
  \hline 
 $ \phantom{{\bar{ \bar  {\bar  {\yng(1,1)}}}}}\!  \! \!\! \! \!  \!\!\!$  {\rm IR}&   $ B^{[A_1 B_2], m}$      &   $ 2   \cdot     {\bar  {\yng(1,1)}}   $     &   $ \frac{N(N-1)}{2}  \cdot   {{\yng (1)}}$      &    $     N(N-1) \cdot  (\cdot) $       &    $ -1 $   \\
 & $ B^{[A_1 B_3], m}$      &   $  (12-N)    \cdot     {\bar {\yng(1)}}   $     &   $ N (12-N)   \cdot   (\cdot)$      &    $   N\cdot  {\yng(1)}$       &    $ -1 $   \\
\hline
\end{tabular}  
  \caption{\footnotesize  $SU(2)$ symmetric phase in the $(2,1)$ model.   
  $A_1, B_1$  stand for  the flavor  indices up to  $N$;     $A_2, B_2$ from $N+1$ to $2N$,  $A_3, B_3$  the last $12-N$.  
  The anomaly matching fails in this case.  }\label{decomposition}
\end{table}
 The candidate baryons have the form, 
 \be   B^{A B, m}=     \psi^{\{ij\},m}  \eta_{i}^A \eta_{j}^B \;,
 \ee
 but it is not possible to achieve the anomaly matching. 

\subsection{Dynamical Abelianization}
 
Assuming that the adjoint condensate forms
\beq  \langle  \psi^{\{ij\},\, 1} \chi_{[ik]}  
\rangle = c^j \, \Lambda^3\, \delta^j_k  \,, \qquad j,k =1, 2, \ldots , N    \ , \label{adjcond} 
\eeq
with $c^j$'s all different
  the Cartan subgroup of $SU(N)_{\rm c}$  survives in the infrared.  
 $SU(2)_{\rm f}$  is broken.    There is a $U(1)$ symmetry which remains unbroken, ${\tilde U}(1)$,   
under which
\be    \psi:  N+12 \;;\qquad   \chi:  - (N+12)  \;;\qquad \eta:    - (N+6) \;.
\ee
The unbroken symmetry group is
\be      SU(N+12)_{\rm f} \times   {\tilde U}(1)\;.
\ee

The low energy degrees of freedom are the fermion fields  $\eta^B_j$ which are unconfined and 
are  weakly coupled to the $U(1)^{N-1} $ photons, the diagonal $\psi^{\{ii\},\, 1}$ and all of   $\psi^{\{ij\},\, 2}$.  
Also there are $3 +1=4$  NG bosons. 

The anomaly equalities for  $SU(12+N)_{\rm f}^3$, $  {\tilde U}(1)  SU(12+N)_{\rm f}^2 $, $ {\tilde U}(1)^3$,  $ {\tilde U}(1)$ can be straightforwardly checked, see Table~\ref{Abelianta}.
\begin{table}[h!t]
  \centering 
  \begin{tabular}{|c|c|c |c|  }
\hline
$ \phantom{{{   {  {\yng(1)}}}}}\!  \! \! \! \! \!\!\!$   & fields          &  $ SU(N+12)$     &   $ {\tilde U}(1)   $  \\
 \hline
  \phantom{\huge i}$ \! \!\!\!\!$ {\rm UV}& $\psi$      &    $  2\cdot  \frac{N(N+1)}{2} \cdot  (\cdot) $    & $N+12$    \\
 &   $\chi$      &    $   \frac{N(N-1)}{2} \cdot  (\cdot) $    & $ -(N+12)  $        \\
 &$ \eta^{A}$      &   $ N \, \cdot  \, {\yng(1)}  $     &   $  -(N+6) $ \\
   \hline 
   \phantom{\huge i}$ \! \!\!\!\!$ \! {\rm IR}&      $  \psi^{ii\;, 1}  $      &  $ N \cdot ( \cdot)   $        &    $  N +12 $   \\
         &       $  \psi^{ij\;, 2}  $      &  $ \frac{N(N+1)}{2} \cdot ( \cdot)   $        &    $  N +12 $   \\
  &     $ \psi \chi \eta^A  \sim   \eta^{A} $      &  $ N \, \cdot  \, {\yng(1)}  $        &    $  -(N+6)  $   \\
\hline
\end{tabular}
  \caption{\footnotesize  The decomposition of the fields in the $(2,1)$  model, assuming the complete dynamical Abelianization. 
  \label{Abelianta} }
  \end{table}

\section{$(N_{\psi}, N_{\chi})=  (1,-1)$}
\label{dieci}

Consider now a model with 
\be   \psi^{\{ij\}}\;, \qquad  {\tilde \chi}^{[ij]}\;, \qquad    \eta_i^A\;,\qquad  A=1,2,\ldots 2N\;,  
\ee
or
\be  
     \,  \yng(2) +  {{\yng(1,1)}} +   2N  \, {\bar {\yng(1)}}\;,
\ee 
i.e.,  a symmetric tensor,  an antisymmetric tensor and $2N$ anti-fundamental multiplets  of $SU(N)$. 
The first coefficient of the beta function is
\be   b_0 = \frac 13\left[  11N-  (  N+2) - (  N-2)  - 2N  \right] =  \frac{7 N}{3} \;.      \label{beta00}
\ee
The symmetry of the system is
\be   SU(N)_{\rm c}\times   SU(2N)_{\rm f} \times U_1(1)\times U_2(1) 
\ee
times some discrete symmetry.   The $U(1)$ charges are:
\bea   && U_1(1):     \qquad    Q_{\psi} =  \frac{1}{N+2}\;, \qquad  Q_{\tilde {\chi}} = - \frac{1}{N-2}\;, \qquad  Q_{\eta} = 0 \;; \qquad 
\nonumber \\   &&  U_2(1):     \qquad    Q_{\psi} =  \frac{1}{N+2}\;, \qquad  Q_{\tilde {\chi}} =0  \;, \qquad  Q_{\eta} = -   \frac{1}{2N} \;, \qquad 
\eea

Possible baryon states are
\be     B^{AB}=\psi^{\{ij\}}  \eta_i^A   \eta_j^B\;, \qquad    {\hat B}^{AB}= {\tilde \chi}^{[ij]}  \eta_i^A   \eta_j^B\;,
\ee
both of which could form either symmetric or antisymmetric tensors in the flavor.  
Confinement without chiral symmetry breaking appears excluded:  there is no way  $ B^{AB}$ or $ {\hat B}^{AB}$ can 
match the UV $SU(2N)_{\rm f}$ anomaly,   $N$. 

\subsection{Color-flavor locking  \label{sec:CF1m1}}

Let us try a color-flavor locking
\bea  && \brc    \psi^{\{ij\}}  \eta_j^A  \ckt = c  \Lambda^3\,\delta^{iA} \;,  \qquad i, A= 1,2,\ldots, N\;,
\nonumber \\ && \ \brc    {\tilde \chi}^{[ij]}  \eta_j^A  \ckt =c' \Lambda^3\delta^{iA}\;,  \qquad  i, A= 1,2,\ldots, N\;.
\eea
The symmetry is broken to 
\be 
 SU(N)_{\rm cf} \times  SU(N)_{\rm f} \times {\tilde U}(1)
\ee
where 
${\tilde U}(1)$ is an unbroken combination of $U_{1,2}(1)$, with charges,
  \be  {\tilde U}(1)    : \qquad     Q_{\psi} =  - 1\;, \qquad  Q_{\tilde {\chi}} =- 1   \;, \qquad  Q_{\eta} =  1 \;.
\ee
Again we list the fields and their decomposition in the low-energy symmetry groups. Assuming that 
the only massless baryons are   $B^{AB}$, with  $A\le N$,  $B\ge N$,  the anomaly matching is obvious, see Table~\ref{CF1m1}.  
 \begin{table}[h!t]
  \centering 
  \begin{tabular}{|c|c|c |c|c|  }
\hline
$ \phantom{{{   {  {\yng(1)}}}}}\!  \! \! \! \! \!\!\!$   & fields   &  $SU(N)_{\rm cf} $    &  $ SU(N)_{\rm f}$    &   $ {\tilde U}(1)   $   \\
\hline
  \phantom{\huge i}$ \! \!\!\!\!$  {\rm UV}&  $\psi$   &   $    { \yng(2)} $  &    $  \frac{N(N+1)}{2}  \cdot (\cdot) $    &   $   - 1 $    \\
 &   $\chi$   &   $    { \yng(1,1)} $  &    $  \frac{N(N-1)}{2}  \cdot (\cdot) $    &   $ - 1$      \\
 &$ \eta^{A_1}$      &   $    {\bar  {\yng(2)}}  \oplus   {\bar  {\yng(1,1)}}   $ &    $ N^2  \cdot (\cdot) $    & $ 1$      \\
 & $ \eta^{A_2}$      &  $ N   \cdot     {\bar  {\yng(1)}}  $     &   $ N \cdot  {\yng(1)}$      &    $ 1$      \\
  \hline  
   $ \phantom{{\bar{ \bar  {\bar  {\yng(1)}}}}}\!  \! \! \! \! \! \!\!\!$  {\rm IR}&   $ B^{A_1 B_2}$      &   $ N  \cdot     {\bar  {\yng(1)}}  $     &   $  N \cdot    {\yng(1)}$      &    $1$          \\
\hline
\end{tabular}
  \caption{\footnotesize  The color-flavor locking scheme for the $(1,-1)$ model.   The flavor indices $A_1, B_1$ stand for 
  those up to $N$,  $A_2, B_2$ for   $N+1, \ldots, 2N$.
  }\label{CF1m1}
\end{table}

\section{Pion decay constant in chiral theories  \label{sec:Fpi}}
\label{undici}

After these exercises with various $(N_{\psi}, N_{\chi}$) models, it would be useful to try to draw some lessons. One concerns the nature of the Nambu-Goldstone bosons (called ``pions" below symbolically) and  the quantity  analogous to the pion-decay constant in the chiral $SU(2)_L \times SU(2)_R$ QCD.  As we shall see, there is some qualitative difference between the wisdom about the chiral dynamics with light quarks in QCD which is a vector-like theory, and what is to be expected in general  chiral theories.

Consider any global continuous symmetry  $G_{\rm f}$  and the associated conserved current $J_{\mu}$,  the field $\phi$ (elementary or composite) which condenses and break  $G_{\rm f}$, and the field 
${\tilde \phi}$  which is transformed into   $\phi$  by the  $G_{\rm f}$  charge 
\be  Q \equiv  \int d^3x    J_{0}\;, \qquad    [Q,  {\tilde \phi}] =    \phi\;, \qquad \brc \phi \ckt  \ne 0\;.
\ee
Thus
\bqa  && \lim_{q_\mu \to 0 }  i q^{\mu} \int d^4x   \, e^{- i q \cdot x} \brc  0| T\{ J_{\mu}(x) \,  {\tilde \phi}(0) \}  | 0 \ckt =  \nonumber \\
&&=   \lim_{q_\mu \to 0 }  \int d^4x   \, e^{- i q \cdot x}    \de_{\mu}   \brc  0| T\{ J_{\mu}(x) \,  {\tilde \phi}(0) \}  | 0 \ckt =  \nonumber \\
&&=   \int d^3x    \brc  0| [  J_{0}(x),   {\tilde \phi}(0)  ]  | 0 \ckt =    \brc  0| [  Q,   {\tilde \phi}(0) ]  | 0 \ckt =   \brc  0| \phi(0) | 0 \ckt  \ne 0 \;.  \label{chWT}
\eea
This Ward-Takahashi like identity  
implies that the two-point function
\be    \int d^4x   \, e^{- i q \cdot x} \brc  0| T\{ J_{\mu}(x) \,  {\tilde \phi}(0) \}  | 0 \ckt  
\ee
is singular at $q \to 0$.  If the $G_{\rm f}$ symmetry is broken spontaneously such a singularity is due to the 
 massless NG boson, $\pi$, such that 
\be   \brc 0| J_{\mu}(q)  |  \pi \ckt =  i q_{\mu}  F_{\pi}\;, \qquad   \brc \pi | {\tilde \phi} | 0 \ckt \ne 0\;,  
\ee
such that  the two point function behaves as 
\be     q^{\mu} \cdot q_{\mu} \,  \frac{F_{\pi}   \brc \pi | {\tilde \phi} | 0 \ckt  }{q^2} \sim \const 
\ee
at $q \to 0$.

   In the standard $SU(2)_L \times SU(2)_R \to SU(2)_V$ chiral symmetry breaking in QCD, the quarks are   
\be   \psi_L =  \left(\begin{array}{c}u_L \\d_L\end{array}\right)\;, \qquad    \psi_R =  \left(\begin{array}{c}u_R \\d_R\end{array}\right)\;,    \ee
and by taking  
\be      \phi =  {\bar \psi_R} \psi_L +  {\rm h.c.  }  \;,  \qquad   {\tilde \phi} =   {\bar \psi_R} t^b  \psi_L -    {\rm h.c.  }   \;;  \qquad  J^{5, a}_{\mu} =   i  {\bar \psi_L}  {\bar \sigma }_{\mu}  t^a \psi_L - (L\leftrightarrow R)
\ee
\be   t^a= \frac{\tau^a}{2}\;, \qquad a=1,2,3\;. 
\ee
It is believed that the field
\be   \brc \phi \ckt=   \brc  {\bar u_R} u_L + {\bar d_R} d_L   + {\rm h.c.  }   \ckt \sim   - \Lambda^3\;, 
\ee
condenses,  leaving $SU(2)_V$ unbroken;  the axial $SU(2)_A$ is broken.    In the QCD $\Lambda$ is of the same order of the confinement 
mass scale, the dynamically generated mass scale of  QCD, 
  \be \Lambda \sim  200  \,\, {\rm MeV}\,.   \label{mass}  \ee 
The pions are associated with the interpolating field
\be    \pi^a \sim    {\tilde \phi}^a =   {\bar \psi_R} t^a\psi_L-    {\rm h.c.  }  \sim {\bar \psi}_D  \gamma^5 t^a \psi_D 
\ee
(where $\psi_D$ is the Dirac spinors for the quarks).  It is natural to expect that  the pion decay constant,  the amplitude with which the current operator
$J^{5, a}_{\mu} $  produces  the pions from the vacuum, is of the same order of magnitude as $\Lambda$ itself,   
\be      F_{\pi}  \sim \Lambda\;.
\ee
Indeed, the best experimental estimate for $F_{\pi}$ is 
\be  F_{\pi} \sim 130 \,\,{\rm MeV}\;, 
\ee 
cfr. with (\ref{mass}).

Now let us study the case of chiral gauge theories, as those considered in this paper.   
To be concrete, consider the dynamical scenarios, Subsection \ref{possible2} 
in the $(2,0)$ model. 
The symmetry breaking pattern is
\be  SU(N)_{\rm c} \times  SU(2)_{\rm f} \times SU(2N+8)_{\rm f} \times U(1)   \to  SU(N)_{\rm cf} \times  {\tilde U}(1) \times U^{\prime}(1) \times SU(8)\;.\label{thistimeBis}
\ee
The Nambu-Goldstone modes are associated with the breaking 
\be    SU(2)_{\rm f} \times SU(2N+8)_{\rm f}  \to  SU(8)  \times U^{\prime}(1)  \;,
\ee
There are 
\be  3  N^2 + 32 N  + 3     \;
\ee
NG bosons.

To simplify the discussion  let us 
concentrate our attention to the two NG bosons associated with the $SU_{\psi}(2)\to  U^{\prime}(1)$ breaking \footnote{Naturally the same discussion holds for other  $3  N^2 + 32 N  + 1$   NG bosons, but the expressions would become more clumsy. }.  
The $SU_{\rm f}(2)$  current   is 
\be    J^a_{\mu} =  i \, {\bar \psi}^{ij, m}  {\bar \sigma }_{\mu}    \left(\frac{\tau^a}{2}\right)_{mn}  \psi^{ij, n}\;,   
\ee
and  the charges are
\be   Q^a=  \int d^3x  \,   J^a_{0}\;. 
\ee
One can choose 
\be      {\tilde \phi^b} =     \sum_{i,j,k,  B} ( \psi^{\{ij\,, m\}}   \eta_i^B)^*  \left(\frac{\tau^b}{2}\right)_{mn}    \psi^{\{k j\,, n \}}  \eta_k^B\;  \label{xpions}
\ee
in (\ref{chWT}):  so that
\be  \brc  [Q^a,   {\tilde \phi^b} ]  \ckt  =  \delta^{ab}  \brc  \sum_{i,j,k,  B} ( \psi^{\{ij\,, m\}}   \eta_i^B)^*   (\psi^{\{k j\,, m \}}  \eta_k^B)  \; \ckt   \ne 0\;.
\ee
An important issue here is the fact that even though the dynamical gauge and flavor symmetry breaking are (by assumption) determined by the ``dynamical Higgs scalar"  condensates
\bea && \brc  \psi^{\{ij\,, 1\}}   \eta_j^B \ckt = \, c \,\Lambda^3  \delta^{i,\,  B }\;,   \qquad \quad \     j, B=1,2,\dots  N\;,   
\nonumber \\   && \brc  \psi^{\{ij\,, 2\}}   \eta_j^B \ckt = \, c' \,\Lambda^3   \delta^{i,\,  B-N }\;,   \qquad   j=1,2,\dots  N\;, \quad B=N+1,\ldots, 2N \;,   \label{condens1111}
\eea
at some mass scale,  $\Lambda$, the pion interpolating fields appearing in the WT identity  must be gauge invariants
such as   (\ref{xpions}), which are 
necessarily four-fermion composites.   On the other hand,  the ``pion decay constant"  is defined as usual,
\be     \brc   0 |   J_{\mu}^a     | \pi^a \ckt = i q_{\mu}  F_{\pi}\;,\qquad  J_{\mu}^a= i  {\bar \psi}^{ij, m}  {\bar \sigma }_{\mu}    \left(\frac{\tau^a}{2}\right)_{mn}  \psi^{ij, n}\;,
\ee
as the amplitude with which the current operator  produces the NG bosons from the vacuum. 
It is quite possible that the pion decay constant  in chiral theories is  such that 
\be   F_{\pi} \ll   \Lambda\;,   \label{suchthat}
\ee
as the bifermion current operator must produce pions, which are four-fermion composite particles, from the vacuum\footnote{Large $N$ scaling would ruin this hierarchy so $N$ must be kept finite.}. 

Another way of seeing the same question is to think of the pion effective action,   
\be {\cal L} ( {\tilde \phi}^a, \partial_{\mu}  {\tilde \phi}^a) = \frac{1}{2}  \partial_{\mu}   {\tilde \phi}^a\,  \partial^{\mu}  {\tilde \phi}^a + \ldots\;, 
\ee 
in which the interaction  strength among the pions is given by $F_{\pi}$.   The effective action involve eight-fermion, sixteen-fermion, etc.  amplitudes, 
and the result such as (\ref{suchthat}) could well be realized by the complicated strong interaction dynamics.

\section{Discussion} 
\label{dodici}

Let us recapitulate the class of  $(N_{\psi}, N_{\chi})$  models,
analyzed here.   The gauge group is taken to be $SU(N)$. By $N_{\psi}, N_{\chi}$  are indicated the numbers of Weyl fermions  $\psi$ or $\chi$   in the representations
\be        \yng(2)\;, \qquad     {\rm or}    \qquad    {\bar { \yng(1,1)}} \;.   
\ee
Let us take $N_{\psi}\ge 0$.  In the case $N_{\chi}<0$,    $-N_{\chi}$ indicates the number of the fields ${\tilde \chi}$    in the representation
\be    \yng(1,1)
\ee
instead.   The number of the fermions in the antifundamental (or  fundamental) representations $\eta^a$  (or  ${\tilde \eta}^a$) is fixed by the condition that the 
gauge group $SU(N)$ be anomaly free.  Also we restrict the numbers   $N_{\psi}, N_{\chi}$  such that the model is asymptotically free.

 The systems considered here are rather rigid.
 No fermion mass terms can be added in the Lagrangian and this also means that no gauge-invariant bifermion condensates can form.
 They cannot be deformed by addition of any other renormalizable potential terms either, including the topological $\theta F_{\mu \nu} {\tilde F}^{\mu \nu}$ term.
The presence of massless chiral matter fermions means all values of $\theta$ are equivalent to $\theta=0$.
 The vacuum, apart from possible symmetry breaking degeneration,  is expected to be unique.
The system is strongly coupled in the infrared.  Our ignorance about these simple models, 
 after more than a half century of studies of quantum field theories, 
 certainly is  severely hindering our capability of finding any application of them in a physical theory describing Nature.

  In the absence of other theoretical tools, we have insisted in this paper upon trying to find possible useful indications following the standard 't Hooft anomaly matching 
  constraints (for application of some new ideas such as the generalized symmetries and higher-form gauging to these chiral   gauge theories, see \cite{BKetal}). 
The main lesson to be learned is perhaps the fact that color-flavor (or color-flavor-flavor) locking    and dynamical Abelianization, 
in various combinations,  always provides natural ways to solve 
these consistency constraints and to find possible phases of the system.

The strategy we used in paper, for all the models, is summarized as follows. 
First we chose a set of bi-fermions operators that may condense.
Since we do not have a gauge invariant bi-fermion in our theories,  we chose among the  gauge-non-invariant ones, possibly guided by the maximal attractive channel (MAC) criterion.
 Condensations has two important effects:  it breaks part or all of the color symmetry and it breaks part or all of the  flavor symmetry. 
The broken part of the gauge group is dynamically Higgsed.
The unbroken part  confines  or  remains in the IR if it is in the Coulomb phase (as for the dynamical Abelianization).
We then have to look at the anomaly matching conditions. 
The part of the flavor symmetry that is broken by the condensate is saturated by massless NG boson poles.  
For the unbroken part instead, we need to find a set of fermions  in the IR to match the computation in the UV.
We then decompose the UV fermion into direct sum  of representations of the unbroken flavor subgroup that remains unbroken. 
Unlike the UV representation, which is chiral, the IR decomposed representations have in general vectorial subsets.
All the vectorial parts can be removed since they presumably get massive and in any case do not contribute to the 't Hooft anomaly of the unbroken group. 
Other fermions remain in the IR as massless baryons and saturate the 't Hooft anomalies.

The fact that in models $(1,0)$ and $(0,1)$ one can find a set of candidate massless fermions saturating the anomalies of the full unbroken flavor symmetries, seems to be fortuitous, 
rather than being a rule.  In fact,  no analogous set of candidate massless baryons can be found in  other $(2,0)$, $(3,0)$, or $(0,2)$, $(0,3)$ models.  
On the other hand,  the color-flavor breaking (dynamical Higgs) phase of the $(1,0)$ and $(0,1)$ models finds natural generalizations in these more complicated systems.

    In this sense, our proposal shares a common feature with the tumbling scheme, but does not  follow  literally  the MAC    criterion with the multi-scale chains of dynamical gauge symmetry breaking, 
as in the original  proposal \cite{Raby}.  There are a few cases, however,  in which the appearance of hierarchy of mass scales, for reasons entirely different from that in the tumbling mechanism, is rather natural.

The local gauge symmetries can never be ``truly'' spontaneously broken, and any dynamical or elementary Higgs mechanism (including the case of the standard Higgs scalar in the Weinberg-Salam electroweak theory) must be re-interpreted  in a gauge-invariant fashion.\footnote{As explained by 't Hooft, the Higgs VEV
of the form $\brc \phi \ckt   =    v\, {\tiny   \left(\begin{array}{c}1 \\0\end{array}\right)} $ found in all textbooks on the electroweak theory, is just a gauge dependent way of describing the gauge-invariant VEV 
$\brc \phi^{\dagger} \phi \ckt$, so is the statement such as the left hand fermion  being equal to  $  \psi_L={\tiny  \left(\begin{array}{c}\nu_L \\e_L\end{array}\right)} $. }  
What happens in the chiral gauge theories considered here is that  the system produces a bifermion composite states  such as 
\be   \psi(x)\eta(x)\;,   \qquad   \psi(x) \chi(x)\;,  \label{suchas}
\ee
which then act as an effective Higgs scalar field.  As these  ``dynamical"  Higgs fields are still strongly coupled in general,  the way  their condensates
and consequent flavor symmetry breaking is reinterpreted in a gauge invariant fashion  may be more complicated than in  the standard electroweak theory 
where the Higgs scalars are weakly coupled and described by perturbation theory.  The proposed dynamical Higgs mechanism does however have a definite statement 
about the {\it flavor} symmetry breaking:  the latter is described by the condensate of the composite (dynamical) Higgs fields such as above, at the mass scale 
associated with them. 

This brings us to a possibly relevant observation made in Section~\ref{sec:Fpi}.  A study of chiral Ward-Takahashi identities shows that, in contrast to what happens in vector-like gauge theory such as QCD,   the  system might generate a  hierarchy of mass scales, between the mass scale of the condensates of the composite Higgs fields
(\ref{suchas}),  $ ``\Lambda"$,   and the quantity corresponding to the pion decay constant, $ ``F_{\pi}"$.  The latter is the amplitude that the (broken) symmetry current
produces a NG boson ( ``pion") from the vacuum.  The fact that in chiral gauge theories  the current is a two-fermion operator, while the pions 
are in general  four-fermion composites,  in contrast to what happens in the case of axial symmetry breaking in vector-like theories,  could imply a large hierarchy, (\ref{suchthat}). 
Such a possibility appears to be worth further studies,  both from theoretical and phenomenological points of view.

\section*{Acknowledgments}
 We thank A.~Luzio,  M.~Shifman and Y.~Tanizaki for useful discussions. This work  is  supported by the INFN special project grant ``GAST (Gauge and String Theories)".

 \appendix 
 
\section{$a$ theorem and the ACS  criterion  \label{athACS}}

For free theory of bosons and fermions, the $a$ and $c$  coefficients are  given by
\be   a=   \frac{1}{360}  (N_S + \frac{11}{2}  N_{\rm f} + 62 N_V)\;, \qquad  c=   \frac{1}{120}  (N_S + 6 N_{\rm f} + 12 N_V)\;,
\ee
where $N_S$ is the number of scalar particles, $N_{\rm f}$ is the number of Weyl fermions, and $N_V$ is the number of vector bosons.
The $a$-theorem tells 
\be    a_{IR}  \le   a_{UV}\;.
\ee
On the other hand, the   free-energy is  
\be f =  N_B + \frac{7}{4}  N_{\rm f} \;,
\ee
where $N_{\rm f}$ is the number of  the Weyl fermions and $N_B$ is the number of bosons. 
The ACS criterion is that  \cite{ACS,ACSS} 
\be    f_{IR}\le f_{UV}\;.
\ee

For simplicity we shall use $\ta = 360 a$.  For $(N_{\psi},  N_{\chi})$   model,
\be  \ta_{UV}=   62 (N^2-1) +  \frac{33}{4}  N_{\psi} N (3+N) - \frac{11}{4}  N_{\chi}  N (N-7)\;,
\ee
\be   \ta_{IR} = N_S + \frac{11}{2}  N_{\rm f} + 62 N_V\;,
\ee
where $N_V, N_S, N_{\rm f}$ are the number of vector bosons, scalars, and Weyl fermions in the infrared. 
For the ACS  free energy, 
\be   f_{UV}=   2 (N^2-1) +\frac{ 7 N_{\psi}}{8} (N^2 + 3 N + 8) +  \frac{ 7 N_{\chi}}{8} (N^2 - 3 N + 8) \;,
\ee
\be   f_{IR}=    N_B + \frac{7}{4}  N_{\rm f}\ .
\ee
We put those two criteria to the test in Tables \ref{at} and \ref{acs} for the theories and their possible IR phases discussed in the paper.
In all cases the $a$ theorem is satisfied, the ACS criterion fails only for the $(3,0)$ and $(0,3)$ models.

 \begin{table}[h!t]
  \centering 
  \begin{tabular}{|c|c |c| c|   }
\hline
Model       &  $ \ta_{UV}  $     &   $\ta_{IR}    $   &    Status   \\
 \hline
  $(1,1)$    CFL ($N\ge 8$) &    $  \frac{135 N^2}{2} + 44 N - 62$    & $106 N - 538$  &   \Checkmark  \\
     \hline
   $(1,1)$   CFL ($N\le 8$)   &    $  \frac{135 N^2}{2} + 44 N - 62 $    & $ - 2 N^2+ \frac{109 N}{2} +2$  &   \Checkmark   \\
   \hline
    $(1,1)$   Abelianiz.    &    $   \frac{135 N^2}{2} + 44 N - 62 $    & $ \frac{223N}{2} - 61 $  & \Checkmark  \\     \hline
    $(1,0)$    No XSB  &    $   \frac{ 281 N^2+99N - 248}{4} $    & $ \frac{ 11 N^2+77 N +132}{4}$    &  \Checkmark  \\     \hline
       $(1,0)$      &    $   \frac{ 281 N^2+99N - 248}{4} $    & $  \frac{ 11 N^2+109  N +4}{4}  $    &   \Checkmark \\     \hline
 $ (2,0)$  (symm)    &   $ \frac{1}{2}  (157 N^2 + 99  N -  124)  $     &   $ \frac{1}{2}  (   17 N^2 +141 N + 2 )$ &   \Checkmark  \\     \hline
  $ (2,0)$      &   $ \frac{1}{2}  (157 N^2 + 99  N -  124)  $     &   $  \frac{1}{2} (17 N^2 +141 N + 8 )$ &     \Checkmark \\     \hline
  $ (3,0)$      &   $\frac{347 N^2}{4} +   \frac{297 N}{4}  - 62 $     &   $  \frac{65 N^2}{4} +   \frac{519 N}{4}  +1 $  &   \Checkmark   \\      \hline
   $ (0,1)$      &   $ \frac{281 N^2}{4}- \frac{99}{4} -62 $     &   $ \frac{11}{4} (N-3)(N-4) $  &   \Checkmark  \\     \hline
    $ (0,2)$      &   $  \frac{1}{2} ( 157 N^2 - 99 N - 124 ) $     &   $  \frac{1}{2}   ( 17 N^2- 125 N + 228 ) $ &  \Checkmark  \\     \hline
    $ (0,3)$      &   $  \frac{1}{4} (  347 N^2 - 297 N - 248 ) $     &   $  \frac{1}{4}  ( 65 N^2-487 N + 876 ) $  &    \Checkmark  \\     \hline
   $ (2,1)$      &   $ \frac{1}{4} (303 N^2 + 275 N - 248 )$     &   $   \frac{1}{4}  (33  N^2 +  297 N + 16) $   &  \Checkmark   \\     \hline
    $ (1,-1)$      &   $\frac{157N^2}{2} - 62 $     &   $  \frac{15 N^2}{2} + 2 $  &  \Checkmark    \\
\hline
\end{tabular}  
  \caption{The $a$ theorem. }
\label{at}
  \end{table}
  
  \begin{table}[h!t]
  \centering 
  \begin{tabular}{|c|c |c| c|   }
\hline
Model       &  $ f_{UV}  $     &   $ f_{IR}    $   &    Status   \\
 \hline
  $(1,1)$     CFL ($N\ge 8$)  &    $  \frac{15 N^2}{4} + 14 N - 2 $    & $ 4 (4 N-7)
$  &   \Checkmark  \\   \hline
   $(1,1)$   CFL ($N\le 8$)     &    $  \frac{15 N^2}{4} + 14 N - 2$    & $2 + \frac{113 N}{4}- 2 N^2
$  &  \Checkmark   \\   \hline
    $(1,1)$    Abelianiz.    &    $  \frac{15 N^2}{4} + 14 N - 2 $    & $ \frac{71 N}{4} -1
$  &  \Checkmark   \\  \hline
    $(1,0)$   No XSB   &    $  \frac{1}{8}  ( 37  N^2+  63 N  - 16)  $    & $\ \frac{7}{8}  ( N^2 +7 N +12)$    & \Checkmark   \\  \hline
      $(1,0)$      &    $ \frac{1}{8} ( 37  N^2+  63 N  - 16)  $    & $  \frac{1}{8}  ( 7 N^2 +113 N +8) $    & \Checkmark   \\  \hline
 $ (2,0)$  symm    &   $  \frac{1}{4} ( 29  N^2+  63 N  -8)  $     &   $  \frac{1}{4}  (19 N^2 +177  N +4) $ &  \Checkmark    \\   \hline
 $ (2,0)$      &   $  \frac{1}{4} ( 29  N^2+  63 N  -8)  $     &   $   \frac{1}{4}  (19 N^2 +177  N +16)  $ &   (\Checkmark)   \\  \hline
  $ (3,0)$      &   $   \frac{79N^2}{8}  + \frac{189 N}{8}  -2 $     &   $   \frac{85 N^2}{8}  + \frac{723 N}{8}  + 8 $  &   {\color {red}X}   \\  \hline
   $ (0,1)$      &   $\frac{37N^2}{8}-  \frac{63 N}{8} -2  $     &   $  \frac{7}{8} (N-3)(N-4)  $  &   \Checkmark  \\  \hline
    $ (0,2)$      &   $  \frac{1}{4} (29 N^2 - 63  N - 8 ) $     &   $  \frac{1}{8} ( 19 N^2- 145 N -  276 )$   &    (\Checkmark)  \\  \hline
    $ (0,3)$      &   $\frac{1}{8}  ( 79 N^2 - 189  N - 16 )  $     &   $  \frac{1}{8} ( 85 N^2- 659 N + 1212 ) $  &   {\color {red}X}    \\  \hline
   $ (2,1)$      &   $ \frac{1}{8} ( 51 N^2 + 175 N - 16 )  $     &   $    \frac{1}{8} (21 N^2 +   189 N + 32) $   &   \Checkmark    \\  \hline
    $ (1,-1)$      &   $ \frac{29 N^2}{4} - 2 $     &   $ \frac{15 N^2}{4} + 2 $  &   \Checkmark    \\
\hline
\end{tabular}  
  \caption{The ACS Criterion. }
\label{acs}
  \end{table}

\end{document}